\documentclass[aps,10pt,prl,twocolumn,letterpaper,superscriptaddress]{revtex4-1}

\usepackage{amsmath}
\usepackage{epsfig}
\usepackage{graphicx, caption}
\usepackage{latexsym}
\usepackage{amssymb}
\usepackage{color}
\usepackage{amsfonts}
\usepackage{bm}
\usepackage{upgreek}
\usepackage{hyperref}

\usepackage{blindtext}
\begin{document}

%% Notice placement of commas and superscripts and use of &
%% in the author list
\title{Two phase transitions driven by surface electron-doping in WTe$_2$}

\author{Antonio Rossi}
\affiliation{Department of Physics, University of California, Davis, CA 95616, USA}
\affiliation{Advanced Light Source, Lawrence Berkeley National Lab, Berkeley, 94720, USA}

\author{Giacomo Resta}
\affiliation{Department of Physics, University of California, Davis, CA 95616, USA}

\author{Seng Huat Lee}
\affiliation{Materials Research Institute, Penn State University, University Park, PA 16802, USA}

\author{Ronald Dean Redwing}
\affiliation{Materials Research Institute, Penn State University, University Park, PA 16802, USA}

%\author{Kevin Dressler}
%\affiliation{Materials Research Institute, Penn State University, University Park, PA 16802, USA}

\author{Chris Jozwiak}
\affiliation{Advanced Light Source, Lawrence Berkeley National Lab, Berkeley, 94720, USA}

\author{Aaron Bostwick}
\affiliation{Advanced Light Source, Lawrence Berkeley National Lab, Berkeley, 94720, USA}

\author{Eli Rotenberg}
\affiliation{Advanced Light Source, Lawrence Berkeley National Lab, Berkeley, 94720, USA}

\author{Sergey Y. Savrasov}
\affiliation{Department of Physics, University of California, Davis, CA 95616, USA}

\author{I. M. Vishik}
\email[]{ivishik@ucdavis.edu}
\affiliation{Department of Physics, University of California, Davis, CA 95616, USA}

\begin{abstract}

 WTe$_2$ is a multifunctional quantum material exhibiting numerous emergent phases in which tuning of the carrier density plays an important role. Here we demonstrate two non-monotonic changes in the electronic structure of WTe$_2$ upon \textit{in-situ} electron doping.  The first phase transition is interpreted in terms of a shear displacement of the top WTe$_2$ layer, which realizes a local crystal structure not normally found in bulk WTe$_2$.  The second phase transition is associated with stronger interactions between the dopant atoms and the host, both through hybridization and electric field.  These results demonstrate that electron-doping can drive structural and electronics changes in bulk WTe$_2$ with implications for realizing nontrivial band structure changes in heterointerfaces and devices.
\end{abstract}
%\doublespacing
\maketitle
%\newpage

Semimetallic two-dimensional (2D) materials, such as WTe$_2$, are characterized by numerous emergent electronic phenomena including large and non-saturating magnetoresistance \cite{ali_large_2014}, superconductivity  \cite{Pan:PressureDrivenSC_WTe2,Zhu:SCinKIntercalatedWTe2,Fatemi:SC_monolayer_WTe2,Sajadi:gateInducedSC}, and multiple topological phases \cite{Soluyanov:WTe2_type2_wsm,Wang:HOTI_betaPhaseWTe2,Tang:QSHI_monolayer,Qian:QSH_monolayer_WTe2}. They are also highly tunable via numerous perturbations including hydrostatic pressure \cite{Pan:PressureDrivenSC_WTe2}, uniaxial strain \cite{Soluyanov:WTe2_type2_wsm}, gating \cite{Fatemi:SC_monolayer_WTe2,Sajadi:gateInducedSC}, alloying \cite{Chang:arcTunableAlloy}, doping \cite{Kim:OriginStructuralPhaseTransition}, intercalation \cite{Zhu:SCinKIntercalatedWTe2}, THz and optical excitation \cite{sie_ultrafast_2019,Hein:ModeResolvedReciprocalSpaceMapping}, and preparation as monolayer \cite{Tang:QSHI_monolayer,Qian:QSH_monolayer_WTe2}.  Together, this creates an attractive platform for understanding and controlling many-body interactions, for switchable devices, and for realizing emergent heterointerface phenomena. 

Carrier concentration plays a key role in many of these emergent phenomena. This material's large and nonsaturating magnetoresistance \cite{ali_large_2014} has been attributed to perfect compensation of electron and hole pockets in the bulk electronic structure \cite{Pletikosic:electronicStructureBasis_magnetoresistance}, albeit not without controversy \cite{Thirupathaiah:MoTe2_uncompensated,Wang:BreakdownCompensation,Wang2:Imbalance}. As a consequence of this near-balance of electrons and holes, the carrier concentration is also temperature-dependent, yielding a Lifshitz transition $\approx 160K$ \cite{Wu:Lifshitz} where the hole pockets are lost. Additionally, the onset of superconductivity under hydrostatic pressure \cite{Kang:SuppresedMRPressureSC} or potassium intercalation \cite{Zhu:SCinKIntercalatedWTe2} in the bulk or with gating in the monolayer \cite{Fatemi:SC_monolayer_WTe2,Sajadi:gateInducedSC} is also associated with an excess of electron-like carriers. Although electron doping is not predicted to yield a structural phase transition, it is clear that it frequently yields a different electronic ground state than the compensated system.

Here we show that a small amount of surface electron doping in WTe$_2$ can induce a shear displacement in the top layer, producing a crystal structure locally similar to a polytype typically not encountered in ambient conditions. This phase transition is evidenced by pronounced changes in low-energy surface electronic structure with support from first-principles calculations.  A second phase transition at higher doping levels affects higher energy band structure, and is associated both with hybridization with dopant bands and the surface Stark effect. These results highlight the variety of electronic structure changes associated with electron doping in WTe$_2$, with relevance to understanding previously observed phase transitions, devices, and heterointerface phenomena. 

\begin{figure*}[ht]
  \captionsetup{justification=raggedright}
  \includegraphics[width=2.0\columnwidth]{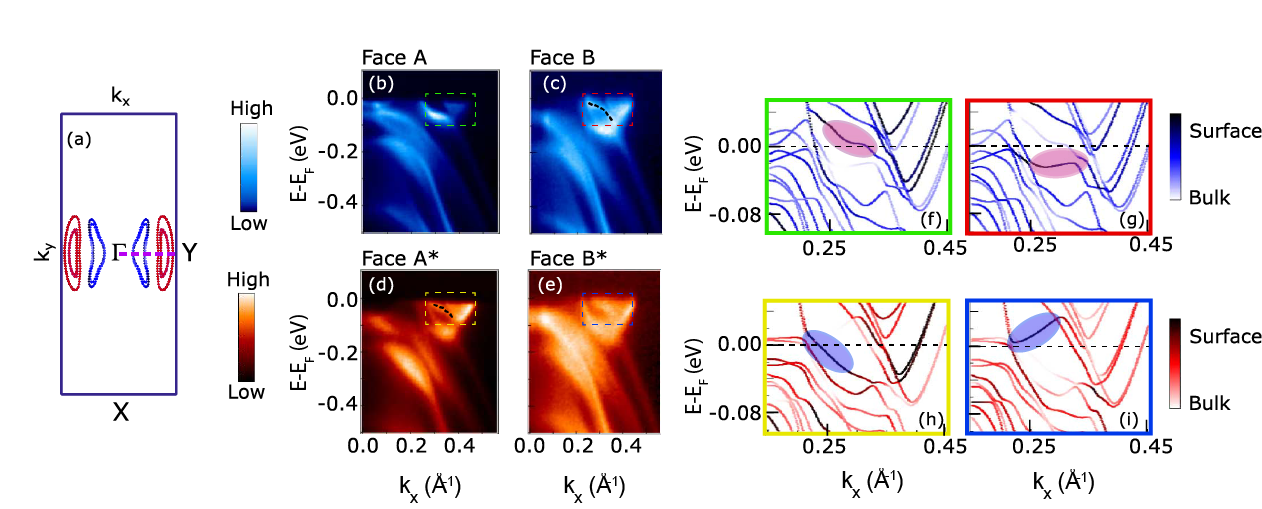}
  \caption{(a) Schematic 2D projection of Brillouin zone (BZ) of WTe$_2$ with bulk electron (red) and hole (blue) Fermi surfaces. Magenta dashed line is cut position for panels (b)-(e). (b)-(c)  ARPES spectra taken on two opposite faces, along $\Gamma$-Y direction, called faces \textquoteleft A\textquoteright and \textquoteleft B\textquoteright, respectively.  (d)-(e) same as (b)-(c) but after K-dosing, called face \textquoteleft A*\textquoteright\ and \textquoteleft B*\textquoteright, respectively.  Dashed curve in panels (c,d) highlights the band which disappears/appears with K-dosing. Panels (f)-(i) show DFT calculations in region of band structure indicated by same color dashed boxes in (b)-(e). (f)-(g) show calculations with $\gamma$ structure, whereas (h)-(i) is with $\beta$.  Color scale indicates bands that are of surface (dark) versus bulk character (light). Pink ovals in (f)-(g) and blue ovals in (h)-(i) highlight changing surface bands. }
  \label{fig:fig1}
\end{figure*}

\begin{figure}[ht]
  \captionsetup{justification=raggedright}
  \includegraphics[width=1\columnwidth]{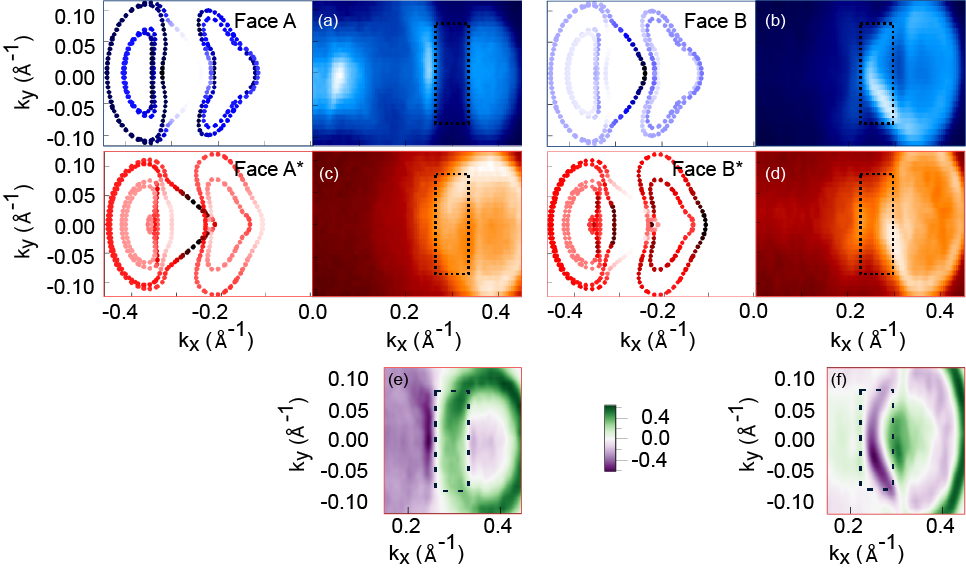}
  \caption{Fermi surface maps with  corresponding DFT calculations mirrored across $k_x=0$, before and after K-dosing.  (a)-(b) before dosing, faces A and B, with corresponding DFT of the $\gamma$ phase mirrored on left. (c)-(d) after K-dosing with DFT of the $\beta$ phase mirrored on the left. (e)-(f) difference spectra (A-A*, B-B*), where green (purple) indicates spectral weight increase (decrease) upon dosing.  In all panels, dashed rectangle marks momentum region of interest where spectral weight increases (decreases) on face A* (B*)}
  \label{fig:fig2}
\end{figure}

\begin{figure}[ht]
\captionsetup{justification=raggedright}
  \includegraphics[width=1.0\columnwidth]{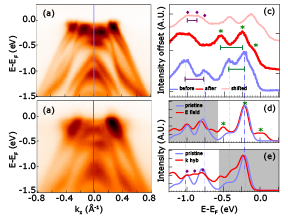}
  \caption{Second phase transition with K-dosing. (a) high symmetry cut along $\Gamma-Y$ before dosing  (b) same cut after large amount of dosing (c) EDC at $\Gamma$ comparing before (blue) and after (red) dosing. Pink EDC: 'after' EDC with 120 meV shift. Curves offset vertically for clarity.  Purple dots and green stars identify comparable features in all panels.  Horizontal bars identify two sets of peaks before and after dosing, high-energy (purple) and low-energy (green). Thin vertical lines: energy position of bands before dosing in experiment and calculations. (d) Slab calculations comparing 'pristine' WTe$_2$ ($\gamma$-phase, blue) to the effect of an electric field of 10 GV/m (red). (e) Slab calculations comparing 'pristine' WTe$_2$ (blue) to the effect of hybridization with a K-overlayer with one K atom per unit cell contributing 0.5 electrons (red). In (d)-(e) EDC broadening is produced by convolving calculated spectra with a gaussian function which best reproduces linewidth of data. }
  \label{fig:fig4}
\end{figure}

In the present angle-resolved photoemission spectroscopy (ARPES) experiments, \textit{in-situ} electron-doping is achieved by depositing variable amounts of potassium on the cleaved WTe$_2$ surface (K-dosing). ARPES experiments were performed at beamline 7.0.2 (MAESTRO) at the Advanced Light Source (ALS). The beam-spot size was $\approx 40$ $\mu m$ for the photon energies of 20 eV (Figs. \ref{fig:fig1},\ref{fig:fig2}) and 90 eV (Fig. \ref{fig:fig4}).  First-principle density function theory calculations were performed using the
localized pseudo-atomic orbital (PAO) method as implemented in
OpenMX~\cite{openmx_pao,openmx}.
High-quality bulk single crystals of WTe$_2$ were synthesized by chemical vapor transport with bromine as the transport agent. More details about experiment and computation can be found in the Supplemental Materials (SM) \cite{Rossi:SM}.

WTe$_2$ is a layered van der Waal material consisting of stacked 2D sheets, and at ambient pressures the alignment between layers leads to a larger simple orthorhombic unit cell containing four formula units, called the $\gamma$ or T$_d$ phase \cite{augustin2000electronic,Dawson:crystalStructure,Kim:OriginStructuralPhaseTransition}. This crystal structure breaks inversion symmetry along the \textit{c}-axis, such that the average Te distance from the W plane is different above and below. This gives rise to different surface bands on opposing sides of the material \cite{bruno2016observation}.  The different surfaces can be accessed by flipping a crystal over, and due to stacking faults, they can also be accessed with subsequent cleaves of a single orientation or different positions on an inhomogeneously cleaved surface \cite{bruno2016observation}.  The latter circumstance necessitates a sufficiently small ARPES beam spot to attain spectra on a single termination.  Fig. \ref{fig:fig1}(b)-(c) highlights the two types of spectra present when different surfaces are exposed: distinct bands are visible near $k_x=0$ and $k_x=0.3$, and in this work, we focus on the latter momentum region.  Following the convention of Ref. \onlinecite{bruno2016observation}, these surfaces are called face \textquoteleft A\textquoteright\ and type \textquoteleft B\textquoteright.

\begin{figure}[ht]
\captionsetup{justification=raggedright}
  \includegraphics[width=1\columnwidth]{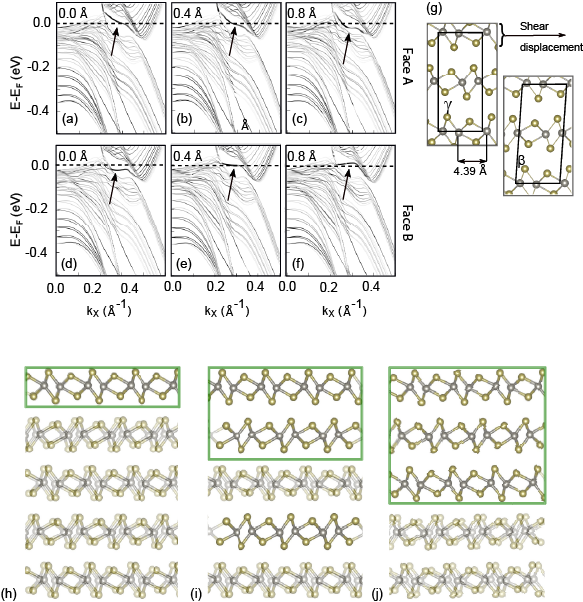}

 \caption{(a)-(c) Slab calculations with different shear displacements of top layer for face \textquoteleft A\textquoteright\ termination.  Darker (lighter) grey denotes more surface-like (bulk-like) character.  Arrows point to surface-derived band to highlight changes as a function of shear displacement. (d)-(f) same for face \textquoteleft B\textquoteright\. (g) definition of shear displacement in $\gamma$ unit cell, together with $\beta$ unit cell (h)-(j) Overlay of metastable crystal structure (see text) with other crystal structures with region of qualitative similarity indicated by a green box. (h) overlay with $\gamma$. (i) overlay with inverted-$\gamma$, with opposite termination. (j) overlay with $\beta$ structure.}
  \label{fig:fig3}
\end{figure}

Fig. \ref{fig:fig1}(b)-(e) shows how \textit{in-situ} K-dosing induces localized changes in the low-energy electronic structure on \textit{both} non-equivalent crystalline faces of WTe$_2$, along a high-symmetry cut indicated in \ref{fig:fig1}(a).  The threshold K-dosing for this to occur corresponds to a shift in the chemical potential of around 0.03 eV or an electron doping of 3-4$\%$ \cite{Rossi:SM}.  Along a high symmetry $\Gamma-Y$ cut on face A (Fig. \ref{fig:fig1}(b),(d)), a new band appears between the electron and hole pocket, at $k_x\approx0.3$. The opposite is observed for face B (Fig. \ref{fig:fig1}(c),(e)) where a band disappears at a similar momentum with K-dosing.   The nonequivalent faces of the K-dosed structure are referred as A* and B*, respectively (Fig. \ref{fig:fig1}(d),(e)).  Density functional theory (DFT) calculations (Fig. \ref{fig:fig1}(f)-(i), \ref{fig:fig3}) are used to give a theoretical basis for the observed changes in the band structure, with the $\beta$ (1T') phase capturing the essential changes in the band structure upon dosing.  The changing portions of the calculated band structure are highlighted with shaded ovals.  In the $\gamma$ phase, the band structure of the two faces differs via a surface-dominated band which is located above $E_F$ on face A, hence not observable by ARPES, but below $E_F$ on face B.  The opposite is true for the $\beta$ crystal structure where face A* shows a surface dominated band between the electron and hole pocket below $E_F$, but face B* has this band above $E_F$.  Most of the other bands in Fig. \ref{fig:fig1}(b)-(e) are unchanged by dosing, which is consistent with prior calculations showing that the $\beta$ and $\gamma$ phase yield very similar bulk bands \cite{Kim:OriginStructuralPhaseTransition}. The structural change we propose is a bit more complicated, and comparison to the $\beta$ phase will be justified later.

The appearance/disappearance of bands between the electron and hole pockets is highlighted further via evolving fermiology, shown in Fig. \ref{fig:fig2}.  Prior to K-dosing, both faces show similar-sized electron and hole pockets, with the hole pocket having weaker intensity at this photon energy due to matrix-element effects.  Faces A and B differ by a segment of Fermi surface extending from the electron to the hole pocket, present only on face B (Fig. \ref{fig:fig2}(a)-(b)). Earlier ARPES studies have demonstrated the 2D (surface-like) nature of these states \cite{wu_observation_2016,Sanchez:SurfaceFermiArc,bruno2016observation}. Present DFT calculations, support the surface-dominated nature of the segments prominent on surface B.   These surface states are sometimes misidentified as the topological Fermi arcs in literature, but they do not connect surface projections of Weyl points and are therefore topologically trivial, though perhaps not irrelevant \cite{Wang:HOTI_betaPhaseWTe2}.  After K-dosing, these surface states on surface B* are no longer observed (Fig. \ref{fig:fig2}(d)). The opposite is seen on face A, where initially only electron and hole pockets are observed, but after dosing, new bands appears between them (Fig. \ref{fig:fig2}(c)), which DFT calculations for the $\beta$ phase indicate to be surface states. This phenomenon on face A/A* was previously reported, but not explained \cite{Zhang:Kdosing_2017}. The doping-induced change is highlighted by subtracting normalized spectra from one another (A-A*, B-B*) in Fig. \ref{fig:fig2}(e)-(f), where states near $k_x=0.3$ are shown to appear (disappear) on face A (B). Notably, while elevated temperature also leads to the diminishment of the hole-like bands, associated with a Lifshitz transition, it does not appear to yield comparable changes in surface bands between the electron and hole pockets\cite{Zhang:Kdosing_2017,Wang:ObservationFermiArcConnection,Thirupathaiah:TempIndeptBandStructWTe2}.  %A shift in chemical potential alone would produce larger electron pockets and smaller hole pockets, as is qualitatively seen on surface A (Fig. \ref{fig:fig2}(e).  On surface B, the topological nature of the phase transition is highlighted via the disappearance of the Fermi arc (Fig. \ref{fig:fig2}(f))

Upon subsequent dosing cycles, the system undergoes a \textit{second} phase transition, shown in Fig \ref{fig:fig4}, corresponding to a chemical potential shift of $\approx 100-130$ meV. The energy separation between bands changes significantly, particularly at the $\Gamma$ point (Fig. \ref{fig:fig4}(c)).  Four bands are resolved between 0 and 1.25 eV binding energy ($E_B$), two at lower $E_B$ and two at higher $E_B$.  Initially, the two bands at lower $E_B$ have an energy separation ($\Delta$) of $\approx 220 meV$ and after, they have $\Delta \approx 300 meV$ (green bars).  Additionally, a new shoulder appears at $E_B\approx 80 meV$.  For the two bands resolved at higher $E_B$, initially $\Delta\approx 300 meV$ and afterwards, $\Delta\approx 170 meV$ (purple bars).  The second phase transition is also seen in off-high-symmetry cuts, shown in the supplementary materials \cite{Rossi:SM}, and these data clarify how trivial effects of K-dosing (rigid band shift and disorder broadening) differ from the present results. Other sudden changes associated with the second phase transition are seen in the W and K core levels, and in the shape of constant-energy maps at equivalent energies in the band structure \cite{Rossi:SM}. The spectral changes at the $\Gamma$ point for the second phase transition are qualitatively compared to the effects of an electric field (Fig. \ref{fig:fig3}(d))  and also to the effects of hybridization with a K-overlayer (Fig. \ref{fig:fig3}(e)) with qualitative agreement in different energy ranges.  

The first phase transition manifests in changes in surface bands at low binding energy.  Below, we will argue that this result can be reproduced by a shear shift of top WTe$_2$ layer for both terminations.  We will also discuss the connection to the $\beta$ phase which is used in corresponding DFT calculations in Figs. \ref{fig:fig1} and \ref{fig:fig2}, and explain the rationale for considering this structural modification.

The $\gamma$ to $\beta$ transition can be visualized as shear displacements of subsequent layers, and the change in energy when one WTe$_2$ layer is subjected to variable amount of shear displacement is shown in SM \cite{Rossi:SM} for both bulk and slab geometry.  A shear displacement of zero corresponds to the WTe$_2$ $\gamma$ phase.  A shear displacement of $\approx0.4\AA$ corresponds to a geometry where Te atoms on adjacent layers are closest.  For neutral or electron-doped WTe$_2$, repulsion of Te antibonding orbitals leads to this configuration constituting a local maximum in energy and being unfavorable.  Hole-doping has the opposite effect, removing electrons from the Te antibonding orbitals and stabilizing this configuration. A displacement of $\approx0.8\AA$ results in the Te atoms of adjacent layers adopting a locally mirrored configuration as compared to the WTe$_2$ $\gamma$ state.  With electron-doping, a shear displacement of $\approx0.8\AA$ corresponds to a \textit{local} minimum in energy, and is referred to as a ‘metastable configuration,’ while the global minimum is found at $<-0.2\AA$, very close to the $\gamma$ configuration.  

Although the metastable configuration, is not a global minimum in energy, the energy difference from the $\gamma$ geometry is very small, $<2 meV$ for one electron per unit cell.  It should be noted that the DFT calculations cannot provide a reliable comparison of such small energy differences between the configurations, especially considering the difficulties in modeling the van der Waals interaction, such that it is possible that the metastable configuration actually has lower energy in the real system.  Additionally, the electric field introduced by the ionization of the K atoms can also favor this shear displacement \cite{Yang:OriginFerroelectricity}.

In Fig. \ref{fig:fig3}(a)-(f), we consider how successive shear displacements of the top WTe$_2$ layer affect band structure in the momentum region of interest.  On surface A, the surface band located between the electron and hole pocket is initially above $E_F$ and hence not observable by ARPES.  It is gradually pushed below $E_F$ by successive shear displacements of the top WTe$_2$ layer along the direction of W-W zigzag chains.  This band is fully below $E_F$ and visible to ARPES between 0.6-0.8$\AA$. Thus the metastable configuration captures this key feature of the data and of the $\beta$-phase electronic structure.  On face B, the opposite is observed, with a surface band initially below $E_F$ between the electron and hole pockets, being pushed above $E_F$ for a shear displacement between 0.6-0.8$\AA$.  A full series of shear displacements is shown in SM \cite{Rossi:SM}.  

We now discuss the similarity between the metastable configuration and other crystal structures (Fig. \ref{fig:fig3}(h)-(j)).  The metastable crystal structure is superimposed on top of the $\gamma$ phase, the inverted $\gamma$ phase (surface A and B reversed), and the $\beta$ phase.  Boxes denote regions of qualitative agreement.  The metastable structure agrees qualitatively with the $\beta$ structure in the first three layers, which explains why the two structures yield qualitatively similar surface bands.  The comparison of ARPES data to the $\beta$ phase calculations is further justified by the surface-sensitivity of the technique.  For photoelectrons with kinetic energy 20 eV, the mean free path is $\approx 6\AA$, such that $\approx99\%$ of the photoemission signal comes from the first three WTe$_2$ layers.   Additionally, the first two layers of the metastable structure agree well with the inverted $\gamma$ structure, giving qualitative support for the observation that the electronic structures on the A and B surfaces seem to ‘swap’ upon electron doping.

We now turn to the second phase transition, achieved with a large amount of K dosing, focusing on the $\Gamma$ point where the largest spectral changes are observed.  Qualitative comparisons are made to two scenarios that may both present: hybridization between WTe$_2$ and the K-overlayer and a dipole electric field produced between the electrons donated to WTe$_2$ and the positively charged K ions at the surface (surface stark effect).  In both cases, the upper limit of physical values is considered, in order to highlight the effects of each phenomenon within the relatively complex band structure of WTe$_2$. The calculations do not consider photoemission matrix element effects and thus do not aim to reproduce the relative heights of peaks. Electric field qualitatively reproduces the results at lower binding energy ($E_B<0.6 eV$, green symbols). Before dosing, two bands are observed, and they move further apart at the second phase transition (green bars) with an additional shoulder developing at lower binding energy. The surface Stark effect results in non-monotonic band shifts because of different orbital characters of different bands, and has previously been demonstrated as a means of band gap engineering in semiconducting 2D materials \cite{Kang:UniversalMechanismBandEngineering,Zhang:InSeBandGapEngineering,Fukutani:ElectricFieldTuning}. Here we indicate that similar effects in a semimetallic 2D material.  At higher binding energy, a hybridization scenario better reproduces the observed effect of the dominant bands moving closer together (purple bars),and possibly exhibiting a third band as a shoulder feature (purple dots). We note that a monotonic chemical potential shift does not capture all observed spectral features (see SM \cite{Rossi:SM}), which is why shifted (unshifted) spectra are used in qualitative comparisons to hybridization (electric field).  These results highlight that multiple phenomena can simultaneously occur at WTe$_2$ interfaces undergping charge transfer, which can either be exploited for band engineering or carefully avoided in heterostructures or devices where pristine bands are desired.

In conclusion, we have demonstrated how surface electron doping can induce two distinct changes in the electronic structure of WTe$_2$. The first transition is interpreted in terms of a metastable shearing of the top WTe$_2$ layer, which yields surface bands qualitatively similar to the $\beta$ phase within the first three layers.  This also demonstrates how surface bands can identify subtle structural changes in WTe$_2$, which only minimally affect the bulk bands.  Additionally, these results suggest that the same shear mode which has been proposed to be activated with hole doping and photoexcitation can also be activated with a very small amount of electron doping. The second phase transition reflects a more complicated interaction with the K atoms and host, involving both hybridization and the surface Stark effect. Both results are broadly relevant to understanding effects of electrostatic gating, interactions with metallic contacts, and heterointerfaces involving charge transfer in devices constructed from semimetallic 2D materials.

\begin{acknowledgments}The authors acknowledge helpful discussions with Dr. Stiven Forti and Dr. Giuseppina Conti.  I.V. and A. R. acknowledge support from UC Davis Startup funds and the Alfred P. Sloan Foundation (FG-2019-12170).  Samples were provided by The Pennsylvania State University Two-Dimensional Crystal Consortium – Materials Innovation Platform (2DCC-MIP) which is supported by NSF cooperative agreement DMR-1539916.  This research used resources of the Advanced Light Source, which is a US Department of Energy Office of Science User Facility under contract no. DE-AC02-05CH11231. G.R. and S.Y.S. were supported by National Science Foundation Grant DMR-1832728.
\end{acknowledgments}

\bibliography{Final_WTe2_v8_PhysRev.bib}

%merlin.mbs apsrev4-1.bst 2010-07-25 4.21a (PWD, AO, DPC) hacked
%Control: key (0)
%Control: author (8) initials jnrlst
%Control: editor formatted (1) identically to author
%Control: production of article title (-1) disabled
%Control: page (0) single
%Control: year (1) truncated
%Control: production of eprint (0) enabled
\begin{thebibliography}{34}%
\makeatletter
\providecommand \@ifxundefined [1]{%
 \@ifx{#1\undefined}
}%
\providecommand \@ifnum [1]{%
 \ifnum #1\expandafter \@firstoftwo
 \else \expandafter \@secondoftwo
 \fi
}%
\providecommand \@ifx [1]{%
 \ifx #1\expandafter \@firstoftwo
 \else \expandafter \@secondoftwo
 \fi
}%
\providecommand \natexlab [1]{#1}%
\providecommand \enquote  [1]{``#1''}%
\providecommand \bibnamefont  [1]{#1}%
\providecommand \bibfnamefont [1]{#1}%
\providecommand \citenamefont [1]{#1}%
\providecommand \href@noop [0]{\@secondoftwo}%
\providecommand \href [0]{\begingroup \@sanitize@url \@href}%
\providecommand \@href[1]{\@@startlink{#1}\@@href}%
\providecommand \@@href[1]{\endgroup#1\@@endlink}%
\providecommand \@sanitize@url [0]{\catcode `\\12\catcode `\$12\catcode
  `\&12\catcode `\#12\catcode `\^12\catcode `\_12\catcode `\%12\relax}%
\providecommand \@@startlink[1]{}%
\providecommand \@@endlink[0]{}%
\providecommand \url  [0]{\begingroup\@sanitize@url \@url }%
\providecommand \@url [1]{\endgroup\@href {#1}{\urlprefix }}%
\providecommand \urlprefix  [0]{URL }%
\providecommand \Eprint [0]{\href }%
\providecommand \doibase [0]{http://dx.doi.org/}%
\providecommand \selectlanguage [0]{\@gobble}%
\providecommand \bibinfo  [0]{\@secondoftwo}%
\providecommand \bibfield  [0]{\@secondoftwo}%
\providecommand \translation [1]{[#1]}%
\providecommand \BibitemOpen [0]{}%
\providecommand \bibitemStop [0]{}%
\providecommand \bibitemNoStop [0]{.\EOS\space}%
\providecommand \EOS [0]{\spacefactor3000\relax}%
\providecommand \BibitemShut  [1]{\csname bibitem#1\endcsname}%
\let\auto@bib@innerbib\@empty
%</preamble>
\bibitem [{\citenamefont {Ali}\ \emph {et~al.}(2014)\citenamefont {Ali},
  \citenamefont {Xiong}, \citenamefont {Flynn}, \citenamefont {Tao},
  \citenamefont {Gibson}, \citenamefont {Schoop}, \citenamefont {Liang},
  \citenamefont {Haldolaarachchige}, \citenamefont {Hirschberger},
  \citenamefont {Ong},\ and\ \citenamefont {Cava}}]{ali_large_2014}%
  \BibitemOpen
  \bibfield  {author} {\bibinfo {author} {\bibfnamefont {M.~N.}\ \bibnamefont
  {Ali}}, \bibinfo {author} {\bibfnamefont {J.}~\bibnamefont {Xiong}}, \bibinfo
  {author} {\bibfnamefont {S.}~\bibnamefont {Flynn}}, \bibinfo {author}
  {\bibfnamefont {J.}~\bibnamefont {Tao}}, \bibinfo {author} {\bibfnamefont
  {Q.~D.}\ \bibnamefont {Gibson}}, \bibinfo {author} {\bibfnamefont {L.~M.}\
  \bibnamefont {Schoop}}, \bibinfo {author} {\bibfnamefont {T.}~\bibnamefont
  {Liang}}, \bibinfo {author} {\bibfnamefont {N.}~\bibnamefont
  {Haldolaarachchige}}, \bibinfo {author} {\bibfnamefont {M.}~\bibnamefont
  {Hirschberger}}, \bibinfo {author} {\bibfnamefont {N.~P.}\ \bibnamefont
  {Ong}}, \ and\ \bibinfo {author} {\bibfnamefont {R.~J.}\ \bibnamefont
  {Cava}},\ }\href {\doibase 10.1038/nature13763} {\bibfield  {journal}
  {\bibinfo  {journal} {Nature}\ }\textbf {\bibinfo {volume} {514}},\ \bibinfo
  {pages} {205} (\bibinfo {year} {2014})}\BibitemShut {NoStop}%
\bibitem [{\citenamefont {Pan}\ \emph {et~al.}(2015)\citenamefont {Pan},
  \citenamefont {Chen}, \citenamefont {Liu}, \citenamefont {Feng},
  \citenamefont {Wei}, \citenamefont {Zhou}, \citenamefont {Chi}, \citenamefont
  {Pi}, \citenamefont {Yen}, \citenamefont {Song}, \citenamefont {Wan},
  \citenamefont {Yang}, \citenamefont {Wang}, \citenamefont {Wang},\ and\
  \citenamefont {Zhang}}]{Pan:PressureDrivenSC_WTe2}%
  \BibitemOpen
  \bibfield  {author} {\bibinfo {author} {\bibfnamefont {X.-C.}\ \bibnamefont
  {Pan}}, \bibinfo {author} {\bibfnamefont {X.}~\bibnamefont {Chen}}, \bibinfo
  {author} {\bibfnamefont {H.}~\bibnamefont {Liu}}, \bibinfo {author}
  {\bibfnamefont {Y.}~\bibnamefont {Feng}}, \bibinfo {author} {\bibfnamefont
  {Z.}~\bibnamefont {Wei}}, \bibinfo {author} {\bibfnamefont {Y.}~\bibnamefont
  {Zhou}}, \bibinfo {author} {\bibfnamefont {Z.}~\bibnamefont {Chi}}, \bibinfo
  {author} {\bibfnamefont {L.}~\bibnamefont {Pi}}, \bibinfo {author}
  {\bibfnamefont {F.}~\bibnamefont {Yen}}, \bibinfo {author} {\bibfnamefont
  {F.}~\bibnamefont {Song}}, \bibinfo {author} {\bibfnamefont {X.}~\bibnamefont
  {Wan}}, \bibinfo {author} {\bibfnamefont {Z.}~\bibnamefont {Yang}}, \bibinfo
  {author} {\bibfnamefont {B.}~\bibnamefont {Wang}}, \bibinfo {author}
  {\bibfnamefont {G.}~\bibnamefont {Wang}}, \ and\ \bibinfo {author}
  {\bibfnamefont {Y.}~\bibnamefont {Zhang}},\ }\href {\doibase
  10.1038/ncomms8805
  https://www.nature.com/articles/ncomms8805#supplementary-information}
  {\bibfield  {journal} {\bibinfo  {journal} {Nature Communications}\ }\textbf
  {\bibinfo {volume} {6}},\ \bibinfo {pages} {7805} (\bibinfo {year}
  {2015})}\BibitemShut {NoStop}%
\bibitem [{\citenamefont {Zhu}\ \emph {et~al.}(2018)\citenamefont {Zhu},
  \citenamefont {Li}, \citenamefont {Lv}, \citenamefont {Li}, \citenamefont
  {Zhu}, \citenamefont {Jia}, \citenamefont {Chen}, \citenamefont {Wen},\ and\
  \citenamefont {Li}}]{Zhu:SCinKIntercalatedWTe2}%
  \BibitemOpen
  \bibfield  {author} {\bibinfo {author} {\bibfnamefont {L.}~\bibnamefont
  {Zhu}}, \bibinfo {author} {\bibfnamefont {Q.-Y.}\ \bibnamefont {Li}},
  \bibinfo {author} {\bibfnamefont {Y.-Y.}\ \bibnamefont {Lv}}, \bibinfo
  {author} {\bibfnamefont {S.}~\bibnamefont {Li}}, \bibinfo {author}
  {\bibfnamefont {X.-Y.}\ \bibnamefont {Zhu}}, \bibinfo {author} {\bibfnamefont
  {Z.-Y.}\ \bibnamefont {Jia}}, \bibinfo {author} {\bibfnamefont {Y.~B.}\
  \bibnamefont {Chen}}, \bibinfo {author} {\bibfnamefont {J.}~\bibnamefont
  {Wen}}, \ and\ \bibinfo {author} {\bibfnamefont {S.-C.}\ \bibnamefont {Li}},\
  }\href {\doibase 10.1021/acs.nanolett.8b03180} {\bibfield  {journal}
  {\bibinfo  {journal} {Nano Letters}\ }\textbf {\bibinfo {volume} {18}},\
  \bibinfo {pages} {6585} (\bibinfo {year} {2018})}\BibitemShut {NoStop}%
\bibitem [{\citenamefont {Fatemi}\ \emph {et~al.}(2018)\citenamefont {Fatemi},
  \citenamefont {Wu}, \citenamefont {Cao}, \citenamefont {Bretheau},
  \citenamefont {Gibson}, \citenamefont {Watanabe}, \citenamefont {Taniguchi},
  \citenamefont {Cava},\ and\ \citenamefont
  {Jarillo-Herrero}}]{Fatemi:SC_monolayer_WTe2}%
  \BibitemOpen
  \bibfield  {author} {\bibinfo {author} {\bibfnamefont {V.}~\bibnamefont
  {Fatemi}}, \bibinfo {author} {\bibfnamefont {S.}~\bibnamefont {Wu}}, \bibinfo
  {author} {\bibfnamefont {Y.}~\bibnamefont {Cao}}, \bibinfo {author}
  {\bibfnamefont {L.}~\bibnamefont {Bretheau}}, \bibinfo {author}
  {\bibfnamefont {Q.~D.}\ \bibnamefont {Gibson}}, \bibinfo {author}
  {\bibfnamefont {K.}~\bibnamefont {Watanabe}}, \bibinfo {author}
  {\bibfnamefont {T.}~\bibnamefont {Taniguchi}}, \bibinfo {author}
  {\bibfnamefont {R.~J.}\ \bibnamefont {Cava}}, \ and\ \bibinfo {author}
  {\bibfnamefont {P.}~\bibnamefont {Jarillo-Herrero}},\ }\href {\doibase
  10.1126/science.aar4642 %J Science} {\ \textbf {\bibinfo {volume} {362}},\
  \bibinfo {pages} {926} (\bibinfo {year} {2018})}\BibitemShut {NoStop}%
\bibitem [{\citenamefont {Sajadi}\ \emph {et~al.}(2018)\citenamefont {Sajadi},
  \citenamefont {Palomaki}, \citenamefont {Fei}, \citenamefont {Zhao},
  \citenamefont {Bement}, \citenamefont {Olsen}, \citenamefont {Luescher},
  \citenamefont {Xu}, \citenamefont {Folk},\ and\ \citenamefont
  {Cobden}}]{Sajadi:gateInducedSC}%
  \BibitemOpen
  \bibfield  {author} {\bibinfo {author} {\bibfnamefont {E.}~\bibnamefont
  {Sajadi}}, \bibinfo {author} {\bibfnamefont {T.}~\bibnamefont {Palomaki}},
  \bibinfo {author} {\bibfnamefont {Z.}~\bibnamefont {Fei}}, \bibinfo {author}
  {\bibfnamefont {W.}~\bibnamefont {Zhao}}, \bibinfo {author} {\bibfnamefont
  {P.}~\bibnamefont {Bement}}, \bibinfo {author} {\bibfnamefont
  {C.}~\bibnamefont {Olsen}}, \bibinfo {author} {\bibfnamefont
  {S.}~\bibnamefont {Luescher}}, \bibinfo {author} {\bibfnamefont
  {X.}~\bibnamefont {Xu}}, \bibinfo {author} {\bibfnamefont {J.~A.}\
  \bibnamefont {Folk}}, \ and\ \bibinfo {author} {\bibfnamefont {D.~H.}\
  \bibnamefont {Cobden}},\ }\href {\doibase 10.1126/science.aar4426} {\bibfield
   {journal} {\bibinfo  {journal} {Science}\ }\textbf {\bibinfo {volume}
  {362}},\ \bibinfo {pages} {922} (\bibinfo {year} {2018})}\BibitemShut
  {NoStop}%
\bibitem [{\citenamefont {Soluyanov}\ \emph {et~al.}(2015)\citenamefont
  {Soluyanov}, \citenamefont {Gresch}, \citenamefont {Wang}, \citenamefont
  {Wu}, \citenamefont {Troyer}, \citenamefont {Dai},\ and\ \citenamefont
  {Bernevig}}]{Soluyanov:WTe2_type2_wsm}%
  \BibitemOpen
  \bibfield  {author} {\bibinfo {author} {\bibfnamefont {A.~A.}\ \bibnamefont
  {Soluyanov}}, \bibinfo {author} {\bibfnamefont {D.}~\bibnamefont {Gresch}},
  \bibinfo {author} {\bibfnamefont {Z.}~\bibnamefont {Wang}}, \bibinfo {author}
  {\bibfnamefont {Q.}~\bibnamefont {Wu}}, \bibinfo {author} {\bibfnamefont
  {M.}~\bibnamefont {Troyer}}, \bibinfo {author} {\bibfnamefont
  {X.}~\bibnamefont {Dai}}, \ and\ \bibinfo {author} {\bibfnamefont {B.~A.}\
  \bibnamefont {Bernevig}},\ }\href {\doibase 10.1038/nature15768
  https://www.nature.com/articles/nature15768#supplementary-information}
  {\bibfield  {journal} {\bibinfo  {journal} {Nature}\ }\textbf {\bibinfo
  {volume} {527}},\ \bibinfo {pages} {495} (\bibinfo {year}
  {2015})}\BibitemShut {NoStop}%
\bibitem [{\citenamefont {Wang}\ \emph {et~al.}(2019)\citenamefont {Wang},
  \citenamefont {Wieder}, \citenamefont {Li}, \citenamefont {Yan},\ and\
  \citenamefont {Bernevig}}]{Wang:HOTI_betaPhaseWTe2}%
  \BibitemOpen
  \bibfield  {author} {\bibinfo {author} {\bibfnamefont {Z.}~\bibnamefont
  {Wang}}, \bibinfo {author} {\bibfnamefont {B.~J.}\ \bibnamefont {Wieder}},
  \bibinfo {author} {\bibfnamefont {J.}~\bibnamefont {Li}}, \bibinfo {author}
  {\bibfnamefont {B.}~\bibnamefont {Yan}}, \ and\ \bibinfo {author}
  {\bibfnamefont {B.~A.}\ \bibnamefont {Bernevig}},\ }\href {\doibase
  10.1103/PhysRevLett.123.186401} {\bibfield  {journal} {\bibinfo  {journal}
  {Phys. Rev. Lett.}\ }\textbf {\bibinfo {volume} {123}},\ \bibinfo {pages}
  {186401} (\bibinfo {year} {2019})}\BibitemShut {NoStop}%
\bibitem [{\citenamefont {Tang}\ \emph {et~al.}(2017)\citenamefont {Tang},
  \citenamefont {Zhang}, \citenamefont {Wong}, \citenamefont {Pedramrazi},
  \citenamefont {Tsai}, \citenamefont {Jia}, \citenamefont {Moritz},
  \citenamefont {Claassen}, \citenamefont {Ryu}, \citenamefont {Kahn},
  \citenamefont {Jiang}, \citenamefont {Yan}, \citenamefont {Hashimoto},
  \citenamefont {Lu}, \citenamefont {Moore}, \citenamefont {Hwang},
  \citenamefont {Hwang}, \citenamefont {Hussain}, \citenamefont {Chen},
  \citenamefont {Ugeda}, \citenamefont {Liu}, \citenamefont {Xie},
  \citenamefont {Devereaux}, \citenamefont {Crommie}, \citenamefont {Mo},\ and\
  \citenamefont {Shen}}]{Tang:QSHI_monolayer}%
  \BibitemOpen
  \bibfield  {author} {\bibinfo {author} {\bibfnamefont {S.}~\bibnamefont
  {Tang}}, \bibinfo {author} {\bibfnamefont {C.}~\bibnamefont {Zhang}},
  \bibinfo {author} {\bibfnamefont {D.}~\bibnamefont {Wong}}, \bibinfo {author}
  {\bibfnamefont {Z.}~\bibnamefont {Pedramrazi}}, \bibinfo {author}
  {\bibfnamefont {H.-Z.}\ \bibnamefont {Tsai}}, \bibinfo {author}
  {\bibfnamefont {C.}~\bibnamefont {Jia}}, \bibinfo {author} {\bibfnamefont
  {B.}~\bibnamefont {Moritz}}, \bibinfo {author} {\bibfnamefont
  {M.}~\bibnamefont {Claassen}}, \bibinfo {author} {\bibfnamefont
  {H.}~\bibnamefont {Ryu}}, \bibinfo {author} {\bibfnamefont {S.}~\bibnamefont
  {Kahn}}, \bibinfo {author} {\bibfnamefont {J.}~\bibnamefont {Jiang}},
  \bibinfo {author} {\bibfnamefont {H.}~\bibnamefont {Yan}}, \bibinfo {author}
  {\bibfnamefont {M.}~\bibnamefont {Hashimoto}}, \bibinfo {author}
  {\bibfnamefont {D.}~\bibnamefont {Lu}}, \bibinfo {author} {\bibfnamefont
  {R.~G.}\ \bibnamefont {Moore}}, \bibinfo {author} {\bibfnamefont {C.-C.}\
  \bibnamefont {Hwang}}, \bibinfo {author} {\bibfnamefont {C.}~\bibnamefont
  {Hwang}}, \bibinfo {author} {\bibfnamefont {Z.}~\bibnamefont {Hussain}},
  \bibinfo {author} {\bibfnamefont {Y.}~\bibnamefont {Chen}}, \bibinfo {author}
  {\bibfnamefont {M.~M.}\ \bibnamefont {Ugeda}}, \bibinfo {author}
  {\bibfnamefont {Z.}~\bibnamefont {Liu}}, \bibinfo {author} {\bibfnamefont
  {X.}~\bibnamefont {Xie}}, \bibinfo {author} {\bibfnamefont {T.~P.}\
  \bibnamefont {Devereaux}}, \bibinfo {author} {\bibfnamefont {M.~F.}\
  \bibnamefont {Crommie}}, \bibinfo {author} {\bibfnamefont {S.-K.}\
  \bibnamefont {Mo}}, \ and\ \bibinfo {author} {\bibfnamefont {Z.-X.}\
  \bibnamefont {Shen}},\ }\href {\doibase 10.1038/nphys4174
  http://www.nature.com/nphys/journal/v13/n7/abs/nphys4174.html#supplementary-information}
  {\bibfield  {journal} {\bibinfo  {journal} {Nat Phys}\ }\textbf {\bibinfo
  {volume} {13}},\ \bibinfo {pages} {683} (\bibinfo {year} {2017})}\BibitemShut
  {NoStop}%
\bibitem [{\citenamefont {Qian}\ \emph {et~al.}(2014)\citenamefont {Qian},
  \citenamefont {Liu}, \citenamefont {Fu},\ and\ \citenamefont
  {Li}}]{Qian:QSH_monolayer_WTe2}%
  \BibitemOpen
  \bibfield  {author} {\bibinfo {author} {\bibfnamefont {X.}~\bibnamefont
  {Qian}}, \bibinfo {author} {\bibfnamefont {J.}~\bibnamefont {Liu}}, \bibinfo
  {author} {\bibfnamefont {L.}~\bibnamefont {Fu}}, \ and\ \bibinfo {author}
  {\bibfnamefont {J.}~\bibnamefont {Li}},\ }\href {\doibase
  10.1126/science.1256815} {\bibfield  {journal} {\bibinfo  {journal}
  {Science}\ }\textbf {\bibinfo {volume} {346}},\ \bibinfo {pages} {1344}
  (\bibinfo {year} {2014})}\BibitemShut {NoStop}%
\bibitem [{\citenamefont {Chang}\ \emph {et~al.}(2016)\citenamefont {Chang},
  \citenamefont {Xu}, \citenamefont {Chang}, \citenamefont {Lee}, \citenamefont
  {Huang}, \citenamefont {Wang}, \citenamefont {Bian}, \citenamefont {Zheng},
  \citenamefont {Sanchez}, \citenamefont {Belopolski}, \citenamefont
  {Alidoust}, \citenamefont {Neupane}, \citenamefont {Bansil}, \citenamefont
  {Jeng}, \citenamefont {Lin},\ and\ \citenamefont
  {Zahid~Hasan}}]{Chang:arcTunableAlloy}%
  \BibitemOpen
  \bibfield  {author} {\bibinfo {author} {\bibfnamefont {T.-R.}\ \bibnamefont
  {Chang}}, \bibinfo {author} {\bibfnamefont {S.-Y.}\ \bibnamefont {Xu}},
  \bibinfo {author} {\bibfnamefont {G.}~\bibnamefont {Chang}}, \bibinfo
  {author} {\bibfnamefont {C.-C.}\ \bibnamefont {Lee}}, \bibinfo {author}
  {\bibfnamefont {S.-M.}\ \bibnamefont {Huang}}, \bibinfo {author}
  {\bibfnamefont {B.}~\bibnamefont {Wang}}, \bibinfo {author} {\bibfnamefont
  {G.}~\bibnamefont {Bian}}, \bibinfo {author} {\bibfnamefont {H.}~\bibnamefont
  {Zheng}}, \bibinfo {author} {\bibfnamefont {D.~S.}\ \bibnamefont {Sanchez}},
  \bibinfo {author} {\bibfnamefont {I.}~\bibnamefont {Belopolski}}, \bibinfo
  {author} {\bibfnamefont {N.}~\bibnamefont {Alidoust}}, \bibinfo {author}
  {\bibfnamefont {M.}~\bibnamefont {Neupane}}, \bibinfo {author} {\bibfnamefont
  {A.}~\bibnamefont {Bansil}}, \bibinfo {author} {\bibfnamefont {H.-T.}\
  \bibnamefont {Jeng}}, \bibinfo {author} {\bibfnamefont {H.}~\bibnamefont
  {Lin}}, \ and\ \bibinfo {author} {\bibfnamefont {M.}~\bibnamefont
  {Zahid~Hasan}},\ }\href {\doibase 10.1038/ncomms10639} {\bibfield  {journal}
  {\bibinfo  {journal} {Nature Communications}\ }\textbf {\bibinfo {volume}
  {7}},\ \bibinfo {pages} {10639} (\bibinfo {year} {2016})}\BibitemShut
  {NoStop}%
\bibitem [{\citenamefont {Kim}\ \emph {et~al.}(2017)\citenamefont {Kim},
  \citenamefont {Kang}, \citenamefont {Hamada},\ and\ \citenamefont
  {Son}}]{Kim:OriginStructuralPhaseTransition}%
  \BibitemOpen
  \bibfield  {author} {\bibinfo {author} {\bibfnamefont {H.-J.}\ \bibnamefont
  {Kim}}, \bibinfo {author} {\bibfnamefont {S.-H.}\ \bibnamefont {Kang}},
  \bibinfo {author} {\bibfnamefont {I.}~\bibnamefont {Hamada}}, \ and\ \bibinfo
  {author} {\bibfnamefont {Y.-W.}\ \bibnamefont {Son}},\ }\href {\doibase
  10.1103/PhysRevB.95.180101} {\bibfield  {journal} {\bibinfo  {journal} {Phys.
  Rev. B}\ }\textbf {\bibinfo {volume} {95}},\ \bibinfo {pages} {180101}
  (\bibinfo {year} {2017})}\BibitemShut {NoStop}%
\bibitem [{\citenamefont {Sie}\ \emph {et~al.}(2019)\citenamefont {Sie},
  \citenamefont {Nyby}, \citenamefont {Pemmaraju}, \citenamefont {Park},
  \citenamefont {Shen}, \citenamefont {Yang}, \citenamefont {Hoffmann},
  \citenamefont {Ofori-Okai}, \citenamefont {Li}, \citenamefont {Reid},
  \citenamefont {Weathersby}, \citenamefont {Mannebach}, \citenamefont
  {Finney}, \citenamefont {Rhodes}, \citenamefont {Chenet}, \citenamefont
  {Antony}, \citenamefont {Balicas}, \citenamefont {Hone}, \citenamefont
  {Devereaux}, \citenamefont {Heinz}, \citenamefont {Wang},\ and\ \citenamefont
  {Lindenberg}}]{sie_ultrafast_2019}%
  \BibitemOpen
  \bibfield  {author} {\bibinfo {author} {\bibfnamefont {E.~J.}\ \bibnamefont
  {Sie}}, \bibinfo {author} {\bibfnamefont {C.~M.}\ \bibnamefont {Nyby}},
  \bibinfo {author} {\bibfnamefont {C.~D.}\ \bibnamefont {Pemmaraju}}, \bibinfo
  {author} {\bibfnamefont {S.~J.}\ \bibnamefont {Park}}, \bibinfo {author}
  {\bibfnamefont {X.}~\bibnamefont {Shen}}, \bibinfo {author} {\bibfnamefont
  {J.}~\bibnamefont {Yang}}, \bibinfo {author} {\bibfnamefont {M.~C.}\
  \bibnamefont {Hoffmann}}, \bibinfo {author} {\bibfnamefont {B.~K.}\
  \bibnamefont {Ofori-Okai}}, \bibinfo {author} {\bibfnamefont
  {R.}~\bibnamefont {Li}}, \bibinfo {author} {\bibfnamefont {A.~H.}\
  \bibnamefont {Reid}}, \bibinfo {author} {\bibfnamefont {S.}~\bibnamefont
  {Weathersby}}, \bibinfo {author} {\bibfnamefont {E.}~\bibnamefont
  {Mannebach}}, \bibinfo {author} {\bibfnamefont {N.}~\bibnamefont {Finney}},
  \bibinfo {author} {\bibfnamefont {D.}~\bibnamefont {Rhodes}}, \bibinfo
  {author} {\bibfnamefont {D.}~\bibnamefont {Chenet}}, \bibinfo {author}
  {\bibfnamefont {A.}~\bibnamefont {Antony}}, \bibinfo {author} {\bibfnamefont
  {L.}~\bibnamefont {Balicas}}, \bibinfo {author} {\bibfnamefont
  {J.}~\bibnamefont {Hone}}, \bibinfo {author} {\bibfnamefont {T.~P.}\
  \bibnamefont {Devereaux}}, \bibinfo {author} {\bibfnamefont {T.~F.}\
  \bibnamefont {Heinz}}, \bibinfo {author} {\bibfnamefont {X.}~\bibnamefont
  {Wang}}, \ and\ \bibinfo {author} {\bibfnamefont {A.~M.}\ \bibnamefont
  {Lindenberg}},\ }\href {\doibase 10.1038/s41586-018-0809-4} {\bibfield
  {journal} {\bibinfo  {journal} {Nature}\ }\textbf {\bibinfo {volume} {565}},\
  \bibinfo {pages} {61} (\bibinfo {year} {2019})}\BibitemShut {NoStop}%
\bibitem [{\citenamefont {Hein}\ \emph {et~al.}(2020)\citenamefont {Hein},
  \citenamefont {Jauernik}, \citenamefont {Erk}, \citenamefont {Yang},
  \citenamefont {Qi}, \citenamefont {Sun}, \citenamefont {Felser},\ and\
  \citenamefont {Bauer}}]{Hein:ModeResolvedReciprocalSpaceMapping}%
  \BibitemOpen
  \bibfield  {author} {\bibinfo {author} {\bibfnamefont {P.}~\bibnamefont
  {Hein}}, \bibinfo {author} {\bibfnamefont {S.}~\bibnamefont {Jauernik}},
  \bibinfo {author} {\bibfnamefont {H.}~\bibnamefont {Erk}}, \bibinfo {author}
  {\bibfnamefont {L.}~\bibnamefont {Yang}}, \bibinfo {author} {\bibfnamefont
  {Y.}~\bibnamefont {Qi}}, \bibinfo {author} {\bibfnamefont {Y.}~\bibnamefont
  {Sun}}, \bibinfo {author} {\bibfnamefont {C.}~\bibnamefont {Felser}}, \ and\
  \bibinfo {author} {\bibfnamefont {M.}~\bibnamefont {Bauer}},\ }\href
  {\doibase 10.1038/s41467-020-16076-0} {\bibfield  {journal} {\bibinfo
  {journal} {Nature Communications}\ }\textbf {\bibinfo {volume} {11}},\
  \bibinfo {pages} {2613} (\bibinfo {year} {2020})}\BibitemShut {NoStop}%
\bibitem [{\citenamefont {Pletikosić}\ \emph {et~al.}(2014)\citenamefont
  {Pletikosić}, \citenamefont {Ali}, \citenamefont {Fedorov}, \citenamefont
  {Cava},\ and\ \citenamefont
  {Valla}}]{Pletikosic:electronicStructureBasis_magnetoresistance}%
  \BibitemOpen
  \bibfield  {author} {\bibinfo {author} {\bibfnamefont {I.}~\bibnamefont
  {Pletikosić}}, \bibinfo {author} {\bibfnamefont {M.~N.}\ \bibnamefont
  {Ali}}, \bibinfo {author} {\bibfnamefont {A.~V.}\ \bibnamefont {Fedorov}},
  \bibinfo {author} {\bibfnamefont {R.~J.}\ \bibnamefont {Cava}}, \ and\
  \bibinfo {author} {\bibfnamefont {T.}~\bibnamefont {Valla}},\ }\href
  {\doibase 10.1103/PhysRevLett.113.216601} {\bibfield  {journal} {\bibinfo
  {journal} {Physical Review Letters}\ }\textbf {\bibinfo {volume} {113}},\
  \bibinfo {pages} {216601} (\bibinfo {year} {2014})}\BibitemShut {NoStop}%
\bibitem [{\citenamefont {Thirupathaiah}\ \emph
  {et~al.}(2017{\natexlab{a}})\citenamefont {Thirupathaiah}, \citenamefont
  {Jha}, \citenamefont {Pal}, \citenamefont {Matias}, \citenamefont {Das},
  \citenamefont {Sivakumar}, \citenamefont {Vobornik}, \citenamefont {Plumb},
  \citenamefont {Shi}, \citenamefont {Ribeiro},\ and\ \citenamefont
  {Sarma}}]{Thirupathaiah:MoTe2_uncompensated}%
  \BibitemOpen
  \bibfield  {author} {\bibinfo {author} {\bibfnamefont {S.}~\bibnamefont
  {Thirupathaiah}}, \bibinfo {author} {\bibfnamefont {R.}~\bibnamefont {Jha}},
  \bibinfo {author} {\bibfnamefont {B.}~\bibnamefont {Pal}}, \bibinfo {author}
  {\bibfnamefont {J.~S.}\ \bibnamefont {Matias}}, \bibinfo {author}
  {\bibfnamefont {P.~K.}\ \bibnamefont {Das}}, \bibinfo {author} {\bibfnamefont
  {P.~K.}\ \bibnamefont {Sivakumar}}, \bibinfo {author} {\bibfnamefont
  {I.}~\bibnamefont {Vobornik}}, \bibinfo {author} {\bibfnamefont {N.~C.}\
  \bibnamefont {Plumb}}, \bibinfo {author} {\bibfnamefont {M.}~\bibnamefont
  {Shi}}, \bibinfo {author} {\bibfnamefont {R.~A.}\ \bibnamefont {Ribeiro}}, \
  and\ \bibinfo {author} {\bibfnamefont {D.~D.}\ \bibnamefont {Sarma}},\ }\href
  {\doibase 10.1103/PhysRevB.95.241105} {\bibfield  {journal} {\bibinfo
  {journal} {Physical Review B}\ }\textbf {\bibinfo {volume} {95}},\ \bibinfo
  {pages} {241105} (\bibinfo {year} {2017}{\natexlab{a}})}\BibitemShut
  {NoStop}%
\bibitem [{\citenamefont {Wang}\ \emph
  {et~al.}(2016{\natexlab{a}})\citenamefont {Wang}, \citenamefont {Wang},
  \citenamefont {Reutt-Robey}, \citenamefont {Paglione},\ and\ \citenamefont
  {Fuhrer}}]{Wang:BreakdownCompensation}%
  \BibitemOpen
  \bibfield  {author} {\bibinfo {author} {\bibfnamefont {Y.}~\bibnamefont
  {Wang}}, \bibinfo {author} {\bibfnamefont {K.}~\bibnamefont {Wang}}, \bibinfo
  {author} {\bibfnamefont {J.}~\bibnamefont {Reutt-Robey}}, \bibinfo {author}
  {\bibfnamefont {J.}~\bibnamefont {Paglione}}, \ and\ \bibinfo {author}
  {\bibfnamefont {M.~S.}\ \bibnamefont {Fuhrer}},\ }\href {\doibase
  10.1103/PhysRevB.93.121108} {\bibfield  {journal} {\bibinfo  {journal} {Phys.
  Rev. B}\ }\textbf {\bibinfo {volume} {93}},\ \bibinfo {pages} {121108}
  (\bibinfo {year} {2016}{\natexlab{a}})}\BibitemShut {NoStop}%
\bibitem [{\citenamefont {Wang}\ \emph {et~al.}(2017)\citenamefont {Wang},
  \citenamefont {Zhang}, \citenamefont {Huang}, \citenamefont {Liu},
  \citenamefont {Liang}, \citenamefont {Zhang}, \citenamefont {Shen},
  \citenamefont {Liu}, \citenamefont {Hu}, \citenamefont {Ding}, \citenamefont
  {Liu}, \citenamefont {Hu}, \citenamefont {He}, \citenamefont {Zhao},
  \citenamefont {Yu}, \citenamefont {Hu}, \citenamefont {Wei}, \citenamefont
  {Mao}, \citenamefont {Shi}, \citenamefont {Jia}, \citenamefont {Zhang},
  \citenamefont {Zhang}, \citenamefont {Yang}, \citenamefont {Wang},
  \citenamefont {Peng}, \citenamefont {Xu}, \citenamefont {Chen},\ and\
  \citenamefont {Zhou}}]{Wang2:Imbalance}%
  \BibitemOpen
  \bibfield  {author} {\bibinfo {author} {\bibfnamefont {C.-L.}\ \bibnamefont
  {Wang}}, \bibinfo {author} {\bibfnamefont {Y.}~\bibnamefont {Zhang}},
  \bibinfo {author} {\bibfnamefont {J.-W.}\ \bibnamefont {Huang}}, \bibinfo
  {author} {\bibfnamefont {G.-D.}\ \bibnamefont {Liu}}, \bibinfo {author}
  {\bibfnamefont {A.-J.}\ \bibnamefont {Liang}}, \bibinfo {author}
  {\bibfnamefont {Y.-X.}\ \bibnamefont {Zhang}}, \bibinfo {author}
  {\bibfnamefont {B.}~\bibnamefont {Shen}}, \bibinfo {author} {\bibfnamefont
  {J.}~\bibnamefont {Liu}}, \bibinfo {author} {\bibfnamefont {C.}~\bibnamefont
  {Hu}}, \bibinfo {author} {\bibfnamefont {Y.}~\bibnamefont {Ding}}, \bibinfo
  {author} {\bibfnamefont {D.-F.}\ \bibnamefont {Liu}}, \bibinfo {author}
  {\bibfnamefont {Y.}~\bibnamefont {Hu}}, \bibinfo {author} {\bibfnamefont
  {S.-L.}\ \bibnamefont {He}}, \bibinfo {author} {\bibfnamefont
  {L.}~\bibnamefont {Zhao}}, \bibinfo {author} {\bibfnamefont {L.}~\bibnamefont
  {Yu}}, \bibinfo {author} {\bibfnamefont {J.}~\bibnamefont {Hu}}, \bibinfo
  {author} {\bibfnamefont {J.}~\bibnamefont {Wei}}, \bibinfo {author}
  {\bibfnamefont {Z.-Q.}\ \bibnamefont {Mao}}, \bibinfo {author} {\bibfnamefont
  {Y.-G.}\ \bibnamefont {Shi}}, \bibinfo {author} {\bibfnamefont {X.-W.}\
  \bibnamefont {Jia}}, \bibinfo {author} {\bibfnamefont {F.-F.}\ \bibnamefont
  {Zhang}}, \bibinfo {author} {\bibfnamefont {S.-J.}\ \bibnamefont {Zhang}},
  \bibinfo {author} {\bibfnamefont {F.}~\bibnamefont {Yang}}, \bibinfo {author}
  {\bibfnamefont {Z.-M.}\ \bibnamefont {Wang}}, \bibinfo {author}
  {\bibfnamefont {Q.-J.}\ \bibnamefont {Peng}}, \bibinfo {author}
  {\bibfnamefont {Z.-Y.}\ \bibnamefont {Xu}}, \bibinfo {author} {\bibfnamefont
  {C.-T.}\ \bibnamefont {Chen}}, \ and\ \bibinfo {author} {\bibfnamefont
  {X.-J.}\ \bibnamefont {Zhou}},\ }\href {\doibase
  10.1088/0256-307x/34/9/097305} {\bibfield  {journal} {\bibinfo  {journal}
  {Chinese Physics Letters}\ }\textbf {\bibinfo {volume} {34}},\ \bibinfo
  {pages} {097305} (\bibinfo {year} {2017})}\BibitemShut {NoStop}%
\bibitem [{\citenamefont {Wu}\ \emph {et~al.}(2015)\citenamefont {Wu},
  \citenamefont {Jo}, \citenamefont {Ochi}, \citenamefont {Huang},
  \citenamefont {Mou}, \citenamefont {Bud'ko}, \citenamefont {Canfield},
  \citenamefont {Trivedi}, \citenamefont {Arita},\ and\ \citenamefont
  {Kaminski}}]{Wu:Lifshitz}%
  \BibitemOpen
  \bibfield  {author} {\bibinfo {author} {\bibfnamefont {Y.}~\bibnamefont
  {Wu}}, \bibinfo {author} {\bibfnamefont {N.~H.}\ \bibnamefont {Jo}}, \bibinfo
  {author} {\bibfnamefont {M.}~\bibnamefont {Ochi}}, \bibinfo {author}
  {\bibfnamefont {L.}~\bibnamefont {Huang}}, \bibinfo {author} {\bibfnamefont
  {D.}~\bibnamefont {Mou}}, \bibinfo {author} {\bibfnamefont {S.~L.}\
  \bibnamefont {Bud'ko}}, \bibinfo {author} {\bibfnamefont {P.~C.}\
  \bibnamefont {Canfield}}, \bibinfo {author} {\bibfnamefont {N.}~\bibnamefont
  {Trivedi}}, \bibinfo {author} {\bibfnamefont {R.}~\bibnamefont {Arita}}, \
  and\ \bibinfo {author} {\bibfnamefont {A.}~\bibnamefont {Kaminski}},\ }\href
  {\doibase 10.1103/PhysRevLett.115.166602} {\bibfield  {journal} {\bibinfo
  {journal} {Phys. Rev. Lett.}\ }\textbf {\bibinfo {volume} {115}},\ \bibinfo
  {pages} {166602} (\bibinfo {year} {2015})}\BibitemShut {NoStop}%
\bibitem [{\citenamefont {Kang}\ \emph {et~al.}(2015)\citenamefont {Kang},
  \citenamefont {Zhou}, \citenamefont {Yi}, \citenamefont {Yang}, \citenamefont
  {Guo}, \citenamefont {Shi}, \citenamefont {Zhang}, \citenamefont {Wang},
  \citenamefont {Zhang}, \citenamefont {Jiang}, \citenamefont {Li},
  \citenamefont {Yang}, \citenamefont {Wu}, \citenamefont {Zhang},
  \citenamefont {Sun},\ and\ \citenamefont
  {Zhao}}]{Kang:SuppresedMRPressureSC}%
  \BibitemOpen
  \bibfield  {author} {\bibinfo {author} {\bibfnamefont {D.}~\bibnamefont
  {Kang}}, \bibinfo {author} {\bibfnamefont {Y.}~\bibnamefont {Zhou}}, \bibinfo
  {author} {\bibfnamefont {W.}~\bibnamefont {Yi}}, \bibinfo {author}
  {\bibfnamefont {C.}~\bibnamefont {Yang}}, \bibinfo {author} {\bibfnamefont
  {J.}~\bibnamefont {Guo}}, \bibinfo {author} {\bibfnamefont {Y.}~\bibnamefont
  {Shi}}, \bibinfo {author} {\bibfnamefont {S.}~\bibnamefont {Zhang}}, \bibinfo
  {author} {\bibfnamefont {Z.}~\bibnamefont {Wang}}, \bibinfo {author}
  {\bibfnamefont {C.}~\bibnamefont {Zhang}}, \bibinfo {author} {\bibfnamefont
  {S.}~\bibnamefont {Jiang}}, \bibinfo {author} {\bibfnamefont
  {A.}~\bibnamefont {Li}}, \bibinfo {author} {\bibfnamefont {K.}~\bibnamefont
  {Yang}}, \bibinfo {author} {\bibfnamefont {Q.}~\bibnamefont {Wu}}, \bibinfo
  {author} {\bibfnamefont {G.}~\bibnamefont {Zhang}}, \bibinfo {author}
  {\bibfnamefont {L.}~\bibnamefont {Sun}}, \ and\ \bibinfo {author}
  {\bibfnamefont {Z.}~\bibnamefont {Zhao}},\ }\href {\doibase
  10.1038/ncomms8804} {\bibfield  {journal} {\bibinfo  {journal} {Nature
  Communications}\ }\textbf {\bibinfo {volume} {6}},\ \bibinfo {pages} {7804}
  (\bibinfo {year} {2015})}\BibitemShut {NoStop}%
\bibitem [{\citenamefont {Ozaki}\ and\ \citenamefont
  {Kino}(2004)}]{openmx_pao}%
  \BibitemOpen
  \bibfield  {author} {\bibinfo {author} {\bibfnamefont {T.}~\bibnamefont
  {Ozaki}}\ and\ \bibinfo {author} {\bibfnamefont {H.}~\bibnamefont {Kino}},\
  }\href {\doibase 10.1103/PhysRevB.69.195113} {\bibfield  {journal} {\bibinfo
  {journal} {Phys. Rev. B}\ }\textbf {\bibinfo {volume} {69}},\ \bibinfo
  {pages} {195113} (\bibinfo {year} {2004})}\BibitemShut {NoStop}%
\bibitem [{\citenamefont {Ozaki}\ \emph {et~al.}()\citenamefont {Ozaki},
  \citenamefont {Kino}, \citenamefont {Yu}, \citenamefont {Han}, \citenamefont
  {Kobayashi}, \citenamefont {Ohfuti}, \citenamefont {Ishii}, \citenamefont
  {Ohwaki}, \citenamefont {Weng},\ and\ \citenamefont {Terakura}}]{openmx}%
  \BibitemOpen
  \bibfield  {author} {\bibinfo {author} {\bibfnamefont {T.}~\bibnamefont
  {Ozaki}}, \bibinfo {author} {\bibfnamefont {H.}~\bibnamefont {Kino}},
  \bibinfo {author} {\bibfnamefont {J.}~\bibnamefont {Yu}}, \bibinfo {author}
  {\bibfnamefont {M.~J.}\ \bibnamefont {Han}}, \bibinfo {author} {\bibfnamefont
  {N.}~\bibnamefont {Kobayashi}}, \bibinfo {author} {\bibfnamefont
  {M.}~\bibnamefont {Ohfuti}}, \bibinfo {author} {\bibfnamefont
  {F.}~\bibnamefont {Ishii}}, \bibinfo {author} {\bibfnamefont
  {T.}~\bibnamefont {Ohwaki}}, \bibinfo {author} {\bibfnamefont
  {H.}~\bibnamefont {Weng}}, \ and\ \bibinfo {author} {\bibfnamefont
  {K.}~\bibnamefont {Terakura}},\ }\href {http://www.openmx-square.org/}
  {\enquote {\bibinfo {title} {Open source package for material explorer},}\
  }\BibitemShut {NoStop}%
\bibitem [{Ros()}]{Rossi:SM}%
  \BibitemOpen
  \href@noop {} {\bibinfo  {journal} {See Supplemental Material at [URL will be
  inserted by publisher] for experimental and computational details, additional
  DFT calculations and photoemission data.}\ }\BibitemShut {NoStop}%
\bibitem [{\citenamefont {Augustin}\ \emph {et~al.}(2000)\citenamefont
  {Augustin}, \citenamefont {Eyert}, \citenamefont {B{\"o}ker}, \citenamefont
  {Frentrup}, \citenamefont {Dwelk}, \citenamefont {Janowitz},\ and\
  \citenamefont {Manzke}}]{augustin2000electronic}%
  \BibitemOpen
\bibfield  {journal} {  }\bibfield  {author} {\bibinfo {author} {\bibfnamefont
  {J.}~\bibnamefont {Augustin}}, \bibinfo {author} {\bibfnamefont
  {V.}~\bibnamefont {Eyert}}, \bibinfo {author} {\bibfnamefont
  {T.}~\bibnamefont {B{\"o}ker}}, \bibinfo {author} {\bibfnamefont
  {W.}~\bibnamefont {Frentrup}}, \bibinfo {author} {\bibfnamefont
  {H.}~\bibnamefont {Dwelk}}, \bibinfo {author} {\bibfnamefont
  {C.}~\bibnamefont {Janowitz}}, \ and\ \bibinfo {author} {\bibfnamefont
  {R.}~\bibnamefont {Manzke}},\ }\href@noop {} {\bibfield  {journal} {\bibinfo
  {journal} {Physical Review B}\ }\textbf {\bibinfo {volume} {62}},\ \bibinfo
  {pages} {10812} (\bibinfo {year} {2000})}\BibitemShut {NoStop}%
\bibitem [{\citenamefont {Dawson}\ and\ \citenamefont
  {Bullett}(1987)}]{Dawson:crystalStructure}%
  \BibitemOpen
  \bibfield  {author} {\bibinfo {author} {\bibfnamefont {W.~G.}\ \bibnamefont
  {Dawson}}\ and\ \bibinfo {author} {\bibfnamefont {D.~W.}\ \bibnamefont
  {Bullett}},\ }\href {\doibase 10.1088/0022-3719/20/36/017} {\bibfield
  {journal} {\bibinfo  {journal} {Journal of Physics C: Solid State Physics}\
  }\textbf {\bibinfo {volume} {20}},\ \bibinfo {pages} {6159} (\bibinfo {year}
  {1987})}\BibitemShut {NoStop}%
\bibitem [{\citenamefont {Bruno}\ \emph {et~al.}(2016)\citenamefont {Bruno},
  \citenamefont {Tamai}, \citenamefont {Wu}, \citenamefont {Cucchi},
  \citenamefont {Barreteau}, \citenamefont {De~La~Torre}, \citenamefont
  {Walker}, \citenamefont {Ricc{\`o}}, \citenamefont {Wang}, \citenamefont
  {Kim} \emph {et~al.}}]{bruno2016observation}%
  \BibitemOpen
  \bibfield  {author} {\bibinfo {author} {\bibfnamefont {F.~Y.}\ \bibnamefont
  {Bruno}}, \bibinfo {author} {\bibfnamefont {A.}~\bibnamefont {Tamai}},
  \bibinfo {author} {\bibfnamefont {Q.}~\bibnamefont {Wu}}, \bibinfo {author}
  {\bibfnamefont {I.}~\bibnamefont {Cucchi}}, \bibinfo {author} {\bibfnamefont
  {C.}~\bibnamefont {Barreteau}}, \bibinfo {author} {\bibfnamefont
  {A.}~\bibnamefont {De~La~Torre}}, \bibinfo {author} {\bibfnamefont {S.~M.}\
  \bibnamefont {Walker}}, \bibinfo {author} {\bibfnamefont {S.}~\bibnamefont
  {Ricc{\`o}}}, \bibinfo {author} {\bibfnamefont {Z.}~\bibnamefont {Wang}},
  \bibinfo {author} {\bibfnamefont {T.}~\bibnamefont {Kim}},  \emph {et~al.},\
  }\href@noop {} {\bibfield  {journal} {\bibinfo  {journal} {Physical Review
  B}\ }\textbf {\bibinfo {volume} {94}},\ \bibinfo {pages} {121112} (\bibinfo
  {year} {2016})}\BibitemShut {NoStop}%
\bibitem [{\citenamefont {Wu}\ \emph {et~al.}(2016)\citenamefont {Wu},
  \citenamefont {Mou}, \citenamefont {Jo}, \citenamefont {Sun}, \citenamefont
  {Huang}, \citenamefont {Bud'ko}, \citenamefont {Canfield},\ and\
  \citenamefont {Kaminski}}]{wu_observation_2016}%
  \BibitemOpen
  \bibfield  {author} {\bibinfo {author} {\bibfnamefont {Y.}~\bibnamefont
  {Wu}}, \bibinfo {author} {\bibfnamefont {D.}~\bibnamefont {Mou}}, \bibinfo
  {author} {\bibfnamefont {N.~H.}\ \bibnamefont {Jo}}, \bibinfo {author}
  {\bibfnamefont {K.}~\bibnamefont {Sun}}, \bibinfo {author} {\bibfnamefont
  {L.}~\bibnamefont {Huang}}, \bibinfo {author} {\bibfnamefont {S.~L.}\
  \bibnamefont {Bud'ko}}, \bibinfo {author} {\bibfnamefont {P.~C.}\
  \bibnamefont {Canfield}}, \ and\ \bibinfo {author} {\bibfnamefont
  {A.}~\bibnamefont {Kaminski}},\ }\href {\doibase 10.1103/PhysRevB.94.121113}
  {\bibfield  {journal} {\bibinfo  {journal} {Phys. Rev. B}\ }\textbf {\bibinfo
  {volume} {94}},\ \bibinfo {pages} {121113} (\bibinfo {year}
  {2016})}\BibitemShut {NoStop}%
\bibitem [{\citenamefont {S\'anchez-Barriga}\ \emph {et~al.}(2016)\citenamefont
  {S\'anchez-Barriga}, \citenamefont {Vergniory}, \citenamefont {Evtushinsky},
  \citenamefont {Aguilera}, \citenamefont {Varykhalov}, \citenamefont
  {Bl\"ugel},\ and\ \citenamefont {Rader}}]{Sanchez:SurfaceFermiArc}%
  \BibitemOpen
  \bibfield  {author} {\bibinfo {author} {\bibfnamefont {J.}~\bibnamefont
  {S\'anchez-Barriga}}, \bibinfo {author} {\bibfnamefont {M.~G.}\ \bibnamefont
  {Vergniory}}, \bibinfo {author} {\bibfnamefont {D.}~\bibnamefont
  {Evtushinsky}}, \bibinfo {author} {\bibfnamefont {I.}~\bibnamefont
  {Aguilera}}, \bibinfo {author} {\bibfnamefont {A.}~\bibnamefont
  {Varykhalov}}, \bibinfo {author} {\bibfnamefont {S.}~\bibnamefont
  {Bl\"ugel}}, \ and\ \bibinfo {author} {\bibfnamefont {O.}~\bibnamefont
  {Rader}},\ }\href {\doibase 10.1103/PhysRevB.94.161401} {\bibfield  {journal}
  {\bibinfo  {journal} {Phys. Rev. B}\ }\textbf {\bibinfo {volume} {94}},\
  \bibinfo {pages} {161401} (\bibinfo {year} {2016})}\BibitemShut {NoStop}%
\bibitem [{\citenamefont {Zhang}\ \emph {et~al.}(2017)\citenamefont {Zhang},
  \citenamefont {Liu}, \citenamefont {Sun}, \citenamefont {Yang}, \citenamefont
  {Jiang}, \citenamefont {Mo}, \citenamefont {Hussain}, \citenamefont {Qian},
  \citenamefont {Fu}, \citenamefont {Yao}, \citenamefont {Lu}, \citenamefont
  {Felser}, \citenamefont {Yan}, \citenamefont {Chen},\ and\ \citenamefont
  {Yang}}]{Zhang:Kdosing_2017}%
  \BibitemOpen
  \bibfield  {author} {\bibinfo {author} {\bibfnamefont {Q.}~\bibnamefont
  {Zhang}}, \bibinfo {author} {\bibfnamefont {Z.}~\bibnamefont {Liu}}, \bibinfo
  {author} {\bibfnamefont {Y.}~\bibnamefont {Sun}}, \bibinfo {author}
  {\bibfnamefont {H.}~\bibnamefont {Yang}}, \bibinfo {author} {\bibfnamefont
  {J.}~\bibnamefont {Jiang}}, \bibinfo {author} {\bibfnamefont {S.-K.}\
  \bibnamefont {Mo}}, \bibinfo {author} {\bibfnamefont {Z.}~\bibnamefont
  {Hussain}}, \bibinfo {author} {\bibfnamefont {X.}~\bibnamefont {Qian}},
  \bibinfo {author} {\bibfnamefont {L.}~\bibnamefont {Fu}}, \bibinfo {author}
  {\bibfnamefont {S.}~\bibnamefont {Yao}}, \bibinfo {author} {\bibfnamefont
  {M.}~\bibnamefont {Lu}}, \bibinfo {author} {\bibfnamefont {C.}~\bibnamefont
  {Felser}}, \bibinfo {author} {\bibfnamefont {B.}~\bibnamefont {Yan}},
  \bibinfo {author} {\bibfnamefont {Y.}~\bibnamefont {Chen}}, \ and\ \bibinfo
  {author} {\bibfnamefont {L.}~\bibnamefont {Yang}},\ }\href {\doibase
  10.1002/pssr.201700209} {\bibfield  {journal} {\bibinfo  {journal} {physica
  status solidi (RRL) – Rapid Research Letters}\ }\textbf {\bibinfo {volume}
  {11}},\ \bibinfo {pages} {1700209} (\bibinfo {year} {2017})}\BibitemShut
  {NoStop}%
\bibitem [{\citenamefont {Wang}\ \emph
  {et~al.}(2016{\natexlab{b}})\citenamefont {Wang}, \citenamefont {Zhang},
  \citenamefont {Huang}, \citenamefont {Nie}, \citenamefont {Liu},
  \citenamefont {Liang}, \citenamefont {Zhang}, \citenamefont {Shen},
  \citenamefont {Liu}, \citenamefont {Hu}, \citenamefont {Ding}, \citenamefont
  {Liu}, \citenamefont {Hu}, \citenamefont {He}, \citenamefont {Zhao},
  \citenamefont {Yu}, \citenamefont {Hu}, \citenamefont {Wei}, \citenamefont
  {Mao}, \citenamefont {Shi}, \citenamefont {Jia}, \citenamefont {Zhang},
  \citenamefont {Zhang}, \citenamefont {Yang}, \citenamefont {Wang},
  \citenamefont {Peng}, \citenamefont {Weng}, \citenamefont {Dai},
  \citenamefont {Fang}, \citenamefont {Xu}, \citenamefont {Chen},\ and\
  \citenamefont {Zhou}}]{Wang:ObservationFermiArcConnection}%
  \BibitemOpen
  \bibfield  {author} {\bibinfo {author} {\bibfnamefont {C.}~\bibnamefont
  {Wang}}, \bibinfo {author} {\bibfnamefont {Y.}~\bibnamefont {Zhang}},
  \bibinfo {author} {\bibfnamefont {J.}~\bibnamefont {Huang}}, \bibinfo
  {author} {\bibfnamefont {S.}~\bibnamefont {Nie}}, \bibinfo {author}
  {\bibfnamefont {G.}~\bibnamefont {Liu}}, \bibinfo {author} {\bibfnamefont
  {A.}~\bibnamefont {Liang}}, \bibinfo {author} {\bibfnamefont
  {Y.}~\bibnamefont {Zhang}}, \bibinfo {author} {\bibfnamefont
  {B.}~\bibnamefont {Shen}}, \bibinfo {author} {\bibfnamefont {J.}~\bibnamefont
  {Liu}}, \bibinfo {author} {\bibfnamefont {C.}~\bibnamefont {Hu}}, \bibinfo
  {author} {\bibfnamefont {Y.}~\bibnamefont {Ding}}, \bibinfo {author}
  {\bibfnamefont {D.}~\bibnamefont {Liu}}, \bibinfo {author} {\bibfnamefont
  {Y.}~\bibnamefont {Hu}}, \bibinfo {author} {\bibfnamefont {S.}~\bibnamefont
  {He}}, \bibinfo {author} {\bibfnamefont {L.}~\bibnamefont {Zhao}}, \bibinfo
  {author} {\bibfnamefont {L.}~\bibnamefont {Yu}}, \bibinfo {author}
  {\bibfnamefont {J.}~\bibnamefont {Hu}}, \bibinfo {author} {\bibfnamefont
  {J.}~\bibnamefont {Wei}}, \bibinfo {author} {\bibfnamefont {Z.}~\bibnamefont
  {Mao}}, \bibinfo {author} {\bibfnamefont {Y.}~\bibnamefont {Shi}}, \bibinfo
  {author} {\bibfnamefont {X.}~\bibnamefont {Jia}}, \bibinfo {author}
  {\bibfnamefont {F.}~\bibnamefont {Zhang}}, \bibinfo {author} {\bibfnamefont
  {S.}~\bibnamefont {Zhang}}, \bibinfo {author} {\bibfnamefont
  {F.}~\bibnamefont {Yang}}, \bibinfo {author} {\bibfnamefont {Z.}~\bibnamefont
  {Wang}}, \bibinfo {author} {\bibfnamefont {Q.}~\bibnamefont {Peng}}, \bibinfo
  {author} {\bibfnamefont {H.}~\bibnamefont {Weng}}, \bibinfo {author}
  {\bibfnamefont {X.}~\bibnamefont {Dai}}, \bibinfo {author} {\bibfnamefont
  {Z.}~\bibnamefont {Fang}}, \bibinfo {author} {\bibfnamefont {Z.}~\bibnamefont
  {Xu}}, \bibinfo {author} {\bibfnamefont {C.}~\bibnamefont {Chen}}, \ and\
  \bibinfo {author} {\bibfnamefont {X.~J.}\ \bibnamefont {Zhou}},\ }\href
  {\doibase 10.1103/PhysRevB.94.241119} {\bibfield  {journal} {\bibinfo
  {journal} {Phys. Rev. B}\ }\textbf {\bibinfo {volume} {94}},\ \bibinfo
  {pages} {241119} (\bibinfo {year} {2016}{\natexlab{b}})}\BibitemShut
  {NoStop}%
\bibitem [{\citenamefont {Thirupathaiah}\ \emph
  {et~al.}(2017{\natexlab{b}})\citenamefont {Thirupathaiah}, \citenamefont
  {Jha}, \citenamefont {Pal}, \citenamefont {Matias}, \citenamefont {Das},
  \citenamefont {Vobornik}, \citenamefont {Ribeiro},\ and\ \citenamefont
  {Sarma}}]{Thirupathaiah:TempIndeptBandStructWTe2}%
  \BibitemOpen
  \bibfield  {author} {\bibinfo {author} {\bibfnamefont {S.}~\bibnamefont
  {Thirupathaiah}}, \bibinfo {author} {\bibfnamefont {R.}~\bibnamefont {Jha}},
  \bibinfo {author} {\bibfnamefont {B.}~\bibnamefont {Pal}}, \bibinfo {author}
  {\bibfnamefont {J.~S.}\ \bibnamefont {Matias}}, \bibinfo {author}
  {\bibfnamefont {P.~K.}\ \bibnamefont {Das}}, \bibinfo {author} {\bibfnamefont
  {I.}~\bibnamefont {Vobornik}}, \bibinfo {author} {\bibfnamefont {R.~A.}\
  \bibnamefont {Ribeiro}}, \ and\ \bibinfo {author} {\bibfnamefont {D.~D.}\
  \bibnamefont {Sarma}},\ }\href {\doibase 10.1103/PhysRevB.96.165149}
  {\bibfield  {journal} {\bibinfo  {journal} {Phys. Rev. B}\ }\textbf {\bibinfo
  {volume} {96}},\ \bibinfo {pages} {165149} (\bibinfo {year}
  {2017}{\natexlab{b}})}\BibitemShut {NoStop}%
\bibitem [{\citenamefont {Yang}\ \emph {et~al.}(2018)\citenamefont {Yang},
  \citenamefont {Wu},\ and\ \citenamefont {Li}}]{Yang:OriginFerroelectricity}%
  \BibitemOpen
  \bibfield  {author} {\bibinfo {author} {\bibfnamefont {Q.}~\bibnamefont
  {Yang}}, \bibinfo {author} {\bibfnamefont {M.}~\bibnamefont {Wu}}, \ and\
  \bibinfo {author} {\bibfnamefont {J.}~\bibnamefont {Li}},\ }\href {\doibase
  10.1021/acs.jpclett.8b03654} {\bibfield  {journal} {\bibinfo  {journal} {The
  Journal of Physical Chemistry Letters}\ }\textbf {\bibinfo {volume} {9}},\
  \bibinfo {pages} {7160} (\bibinfo {year} {2018})}\BibitemShut {NoStop}%
\bibitem [{\citenamefont {Kang}\ \emph {et~al.}(2017)\citenamefont {Kang},
  \citenamefont {Kim}, \citenamefont {Ryu}, \citenamefont {Jung}, \citenamefont
  {Kim}, \citenamefont {Moreschini}, \citenamefont {Jozwiak}, \citenamefont
  {Rotenberg}, \citenamefont {Bostwick},\ and\ \citenamefont
  {Kim}}]{Kang:UniversalMechanismBandEngineering}%
  \BibitemOpen
  \bibfield  {author} {\bibinfo {author} {\bibfnamefont {M.}~\bibnamefont
  {Kang}}, \bibinfo {author} {\bibfnamefont {B.}~\bibnamefont {Kim}}, \bibinfo
  {author} {\bibfnamefont {S.~H.}\ \bibnamefont {Ryu}}, \bibinfo {author}
  {\bibfnamefont {S.~W.}\ \bibnamefont {Jung}}, \bibinfo {author}
  {\bibfnamefont {J.}~\bibnamefont {Kim}}, \bibinfo {author} {\bibfnamefont
  {L.}~\bibnamefont {Moreschini}}, \bibinfo {author} {\bibfnamefont
  {C.}~\bibnamefont {Jozwiak}}, \bibinfo {author} {\bibfnamefont
  {E.}~\bibnamefont {Rotenberg}}, \bibinfo {author} {\bibfnamefont
  {A.}~\bibnamefont {Bostwick}}, \ and\ \bibinfo {author} {\bibfnamefont
  {K.~S.}\ \bibnamefont {Kim}},\ }\href {\doibase 10.1021/acs.nanolett.6b04775}
  {\bibfield  {journal} {\bibinfo  {journal} {Nano Letters}\ }\textbf {\bibinfo
  {volume} {17}},\ \bibinfo {pages} {1610} (\bibinfo {year} {2017})},\ \bibinfo
  {note} {pMID: 28118710}\BibitemShut {NoStop}%
\bibitem [{\citenamefont {Zhang}\ \emph {et~al.}(2019)\citenamefont {Zhang},
  \citenamefont {Chen}, \citenamefont {Bouaziz}, \citenamefont {Giorgetti},
  \citenamefont {Yi}, \citenamefont {Avila}, \citenamefont {Tian},
  \citenamefont {Shukla}, \citenamefont {Perfetti}, \citenamefont {Fan},
  \citenamefont {Li},\ and\ \citenamefont
  {Bendounan}}]{Zhang:InSeBandGapEngineering}%
  \BibitemOpen
  \bibfield  {author} {\bibinfo {author} {\bibfnamefont {Z.}~\bibnamefont
  {Zhang}}, \bibinfo {author} {\bibfnamefont {Z.}~\bibnamefont {Chen}},
  \bibinfo {author} {\bibfnamefont {M.}~\bibnamefont {Bouaziz}}, \bibinfo
  {author} {\bibfnamefont {C.}~\bibnamefont {Giorgetti}}, \bibinfo {author}
  {\bibfnamefont {H.}~\bibnamefont {Yi}}, \bibinfo {author} {\bibfnamefont
  {J.}~\bibnamefont {Avila}}, \bibinfo {author} {\bibfnamefont
  {B.}~\bibnamefont {Tian}}, \bibinfo {author} {\bibfnamefont {A.}~\bibnamefont
  {Shukla}}, \bibinfo {author} {\bibfnamefont {L.}~\bibnamefont {Perfetti}},
  \bibinfo {author} {\bibfnamefont {D.}~\bibnamefont {Fan}}, \bibinfo {author}
  {\bibfnamefont {Y.}~\bibnamefont {Li}}, \ and\ \bibinfo {author}
  {\bibfnamefont {A.}~\bibnamefont {Bendounan}},\ }\href {\doibase
  10.1021/acsnano.9b07144} {\bibfield  {journal} {\bibinfo  {journal} {ACS
  Nano}\ }\textbf {\bibinfo {volume} {13}},\ \bibinfo {pages} {13486} (\bibinfo
  {year} {2019})},\ \bibinfo {note} {pMID: 31644265}\BibitemShut {NoStop}%
\bibitem [{\citenamefont {Fukutani}\ \emph {et~al.}(2019)\citenamefont
  {Fukutani}, \citenamefont {Stania}, \citenamefont {Jung}, \citenamefont
  {Schwier}, \citenamefont {Shimada}, \citenamefont {Kwon}, \citenamefont
  {Kim},\ and\ \citenamefont {Yeom}}]{Fukutani:ElectricFieldTuning}%
  \BibitemOpen
  \bibfield  {author} {\bibinfo {author} {\bibfnamefont {K.}~\bibnamefont
  {Fukutani}}, \bibinfo {author} {\bibfnamefont {R.}~\bibnamefont {Stania}},
  \bibinfo {author} {\bibfnamefont {J.}~\bibnamefont {Jung}}, \bibinfo {author}
  {\bibfnamefont {E.~F.}\ \bibnamefont {Schwier}}, \bibinfo {author}
  {\bibfnamefont {K.}~\bibnamefont {Shimada}}, \bibinfo {author} {\bibfnamefont
  {C.~I.}\ \bibnamefont {Kwon}}, \bibinfo {author} {\bibfnamefont {J.~S.}\
  \bibnamefont {Kim}}, \ and\ \bibinfo {author} {\bibfnamefont {H.~W.}\
  \bibnamefont {Yeom}},\ }\href {\doibase 10.1103/PhysRevLett.123.206401}
  {\bibfield  {journal} {\bibinfo  {journal} {Phys. Rev. Lett.}\ }\textbf
  {\bibinfo {volume} {123}},\ \bibinfo {pages} {206401} (\bibinfo {year}
  {2019})}\BibitemShut {NoStop}%
\end{thebibliography}%


\begin{thebibliography}{10}
\expandafter\ifx\csname url\endcsname\relax
  \def\url#1{\texttt{#1}}\fi
\expandafter\ifx\csname urlprefix\endcsname\relax\def\urlprefix{URL }\fi
\providecommand{\bibinfo}[2]{#2}
\providecommand{\eprint}[2][]{\url{#2}}

\bibitem{openmx_pao}
\bibinfo{author}{Ozaki, T.} \& \bibinfo{author}{Kino, H.}
\newblock \bibinfo{title}{Numerical atomic basis orbitals from {H} to {K}r}.
\newblock \emph{\bibinfo{journal}{Phys. Rev. B}} \textbf{\bibinfo{volume}{69}},
  \bibinfo{pages}{195113} (\bibinfo{year}{2004}).
\newblock \urlprefix\url{https://link.aps.org/doi/10.1103/PhysRevB.69.195113}.

\bibitem{openmx}
\bibinfo{author}{Ozaki, T.} \emph{et~al.}
\newblock \bibinfo{title}{Open source package for material explorer}.
\newblock \urlprefix\url{http://www.openmx-square.org/}.

\bibitem{pbeGGA}
\bibinfo{author}{Perdew, J.~P.}, \bibinfo{author}{Burke, K.} \&
  \bibinfo{author}{Ernzerhof, M.}
\newblock \bibinfo{title}{Generalized gradient approximation made simple}.
\newblock \emph{\bibinfo{journal}{Phys. Rev. Lett.}}
  \textbf{\bibinfo{volume}{77}}, \bibinfo{pages}{3865--3868}
  (\bibinfo{year}{1996}).
\newblock \urlprefix\url{https://link.aps.org/doi/10.1103/PhysRevLett.77.3865}.

\bibitem{norm_pseudo}
\bibinfo{author}{Troullier, N.} \& \bibinfo{author}{Martins, J.~L.}
\newblock \bibinfo{title}{Efficient pseudopotentials for plane-wave
  calculations}.
\newblock \emph{\bibinfo{journal}{Phys. Rev. B}} \textbf{\bibinfo{volume}{43}},
  \bibinfo{pages}{1993--2006} (\bibinfo{year}{1991}).
\newblock \urlprefix\url{https://link.aps.org/doi/10.1103/PhysRevB.43.1993}.

\bibitem{sie_ultrafast_2019}
\bibinfo{author}{Sie, E.~J.} \emph{et~al.}
\newblock \bibinfo{title}{An ultrafast symmetry switch in a {Weyl} semimetal}.
\newblock \emph{\bibinfo{journal}{Nature}} \textbf{\bibinfo{volume}{565}},
  \bibinfo{pages}{61--66} (\bibinfo{year}{2019}).
\newblock \urlprefix\url{http://www.nature.com/articles/s41586-018-0809-4}.

\bibitem{Petersson:Kmetal_KPS}
\bibinfo{author}{Petersson, L.-G.} \& \bibinfo{author}{Karlsson, S.-E.}
\newblock \bibinfo{title}{Clean and oxygen exposed potassium studied by
  photoelectron spectroscopy}.
\newblock \emph{\bibinfo{journal}{Physica Scripta}}
  \textbf{\bibinfo{volume}{16}}, \bibinfo{pages}{425--431}
  (\bibinfo{year}{1977}).
\newblock \urlprefix\url{https://doi.org/10.1088%2F0031-8949%2F16%2F5-6%2F041}.

\bibitem{Cardona:CoreLevels}
\bibinfo{author}{Cardona, M.} \& \bibinfo{author}{Ley, L.}
\newblock \bibinfo{title}{Photoemission in solids. vol. 1: General principles;
  vol. 2: Case studies}.
\newblock \emph{\bibinfo{journal}{Topics in Applied Physics, Berlin: Springer,
  1978, edited by Cardona, M.; Ley, L.}}  (\bibinfo{year}{1978}).

\bibitem{Broden:Kdosing_Fe}
\bibinfo{author}{Broden, G.} \& \bibinfo{author}{Bonzel, H.}
\newblock \bibinfo{title}{Potassium adsorption on {F}e (110)}.
\newblock \emph{\bibinfo{journal}{Surface Science}}
  \textbf{\bibinfo{volume}{84}}, \bibinfo{pages}{106--120}
  (\bibinfo{year}{1979}).

\bibitem{Ozawa:K_adsorb}
\bibinfo{author}{Ozawa, K.} \emph{et~al.}
\newblock \bibinfo{title}{Adsorption of {K} on {N}b{C} (100): photoemission and
  thermal desorption study}.
\newblock \emph{\bibinfo{journal}{Surface science}}
  \textbf{\bibinfo{volume}{336}}, \bibinfo{pages}{93--100}
  (\bibinfo{year}{1995}).

\bibitem{Trioni:AlkaliAdsorption_review}
\bibinfo{author}{Trioni, M.}, \bibinfo{author}{Achilli, S.} \&
  \bibinfo{author}{Chulkov, E.}
\newblock \bibinfo{title}{Key ingredients of the alkali atom – metal surface
  interaction: Chemical bonding versus spectral properties}.
\newblock \emph{\bibinfo{journal}{Progress in Surface Science}}
  \textbf{\bibinfo{volume}{88}}, \bibinfo{pages}{160 -- 170}
  (\bibinfo{year}{2013}).
\newblock
  \urlprefix\url{http://www.sciencedirect.com/science/article/pii/S007968161300018X}.

\bibitem{Thirupathaiah:TempIndeptBandStructWTe2}
\bibinfo{author}{Thirupathaiah, S.} \emph{et~al.}
\newblock \bibinfo{title}{Temperature-independent band structure of {W}{T}e$_2$
  as observed from angle-resolved photoemission spectroscopy}.
\newblock \emph{\bibinfo{journal}{Phys. Rev. B}} \textbf{\bibinfo{volume}{96}},
  \bibinfo{pages}{165149} (\bibinfo{year}{2017}).
\newblock \urlprefix\url{https://link.aps.org/doi/10.1103/PhysRevB.96.165149}.

\bibitem{Wang:MonolayerMoTe2PhaseChange}
\bibinfo{author}{Wang, Y.} \emph{et~al.}
\newblock \bibinfo{title}{Structural phase transition in monolayer
  {M}o{T}e$_{2}$ driven by electrostatic doping}.
\newblock \emph{\bibinfo{journal}{Nature}} \textbf{\bibinfo{volume}{550}},
  \bibinfo{pages}{487} (\bibinfo{year}{2017}).
\newblock \urlprefix\url{https://doi.org/10.1038/nature24043}.

\bibitem{OriginsTransitions}
\bibinfo{author}{Kim, H.-J.}, \bibinfo{author}{Kang, S.-H.},
  \bibinfo{author}{Hamada, I.} \& \bibinfo{author}{Son, Y.-W.}
\newblock \bibinfo{title}{Origins of the structural phase transitions in
  {M}o{T}e$_2$ and {W}{T}e$_2$}.
\newblock \emph{\bibinfo{journal}{Phys. Rev. B}} \textbf{\bibinfo{volume}{95}},
  \bibinfo{pages}{180101} (\bibinfo{year}{2017}).
\newblock \urlprefix\url{https://link.aps.org/doi/10.1103/PhysRevB.95.180101}.

\bibitem{WTe2Brown}
\bibinfo{author}{Brown, B.~E.}
\newblock \bibinfo{title}{The crystal structures of {W}{T}e$_2$ and
  high-temperature {M}o{T}e$_2$}.
\newblock \emph{\bibinfo{journal}{Acta Crystallographica}}
  \textbf{\bibinfo{volume}{20}}, \bibinfo{pages}{268--274}
  (\bibinfo{year}{1966}).
\newblock
  \urlprefix\url{https://onlinelibrary.wiley.com/doi/abs/10.1107/S0365110X66000513}.

\bibitem{WTe2_gamma2}
\bibinfo{author}{Mar, A.}, \bibinfo{author}{Jobic, S.} \&
  \bibinfo{author}{Ibers, J.~A.}
\newblock \bibinfo{title}{Metal-metal vs tellurium-tellurium bonding in wte2
  and its ternary variants tairte4 and nbirte4}.
\newblock \emph{\bibinfo{journal}{Journal of the American Chemical Society}}
  \textbf{\bibinfo{volume}{114}}, \bibinfo{pages}{8963--8971}
  (\bibinfo{year}{1992}).
\newblock \urlprefix\url{https://doi.org/10.1021/ja00049a029}.

\bibitem{TunableWeylMoWTe2}
\bibinfo{author}{Chang, T.-R.} \emph{et~al.}
\newblock \bibinfo{title}{Prediction of an arc-tunable weyl fermion metallic
  state in mo$_x$w$_{1-x}$te$_2$}.
\newblock \emph{\bibinfo{journal}{Nature Communications}}
  \textbf{\bibinfo{volume}{7}}, \bibinfo{pages}{10639} (\bibinfo{year}{2016}).
\newblock \urlprefix\url{https://doi.org/10.1038/ncomms10639}.
\newblock \bibinfo{note}{Article}.

\bibitem{WTe2Bernevig}
\bibinfo{author}{Soluyanov, A.~A.} \emph{et~al.}
\newblock \bibinfo{title}{Type-{II} {W}eyl semimetals}.
\newblock \emph{\bibinfo{journal}{Nature}} \textbf{\bibinfo{volume}{527}},
  \bibinfo{pages}{495} (\bibinfo{year}{2015}).
\newblock \urlprefix\url{https://doi.org/10.1038/nature15768}.

\bibitem{ELK}
\bibinfo{title}{The elk code}.
\newblock \urlprefix\url{http://elk.sourceforge.net/}.

\end{thebibliography}
\end{document}

\title{Two phase transitions driven by surface electron-doping in WTe$_2$: Supplementary Materials}

\maketitle
\newpage

\section*{Materials and methods}
\textbf{First-principle calculations}: First-principle density function theory calculations were performed using the
localized pseudo-atomic orbital (PAO) method as implemented in
OpenMX~\cite{openmx_pao,openmx}.
All calculations were conducted using DFT with spin orbit coupling included and where the Perdew-Burke-Ernzerhof
generalized gradient approximation (PBE-GGA)~\cite{pbeGGA} was used for the exchange-correlation energy.
Fully relativistic norm-conserving pseudo-potentials~\cite{norm_pseudo} and
pseudo-atomic orbitals for each atomic type were taken from the 2013 OpenMX database (PBE13).
The following radial cutoffs (in units of Bohrs) and orbital basis were used: 7.0-s3p2d2 for W, 7.0-s3p2d2 for Te, and 12.0-s2p2d2 for K.
This basis was found to provide an optimal convergence for the bulk band structure as confirmed through comparisons with calculations using a larger radial cutoff and orbital basis (9.0-s3p3d3f1 for W and 9.0-s3p3d3f1 for Te).
An energy cutoff of 200 Ry and a k-point sampling of $8 \times 15
\times 3$ was used for bulk calculations while a sampling of $8 \times 15
\times 1$ was used for surface state calculations.

To calculate the surface states in Figs. 1 and 2 of the manuscript, a supercell consisting of a 3-unit-cell (6-layer) thick
(001)-oriented slab structure and a vacuum layer of 12 \AA\ was constructed
for both the WTe$_2$-$\gamma$ and WTe$_2$-$\beta$ phases, which was found to provide reasonable convergence for the surface states.  A 6-unit-cell thick slab was used in Fig. 4 of the manuscript, and in this SM we show that the two yield similar agreement with ARPES data for the bands of interest.

The surface projected band structure was then determined by projecting the wave-functions to the outermost two layers of the material.

\textbf{ARPES}: ARPES experiments were performed at the Microscopic and Electronic Structure Observatory (MAESTRO) at the Advanced Light Source.  Samples were cleaved in glovebox and measured in microARPES end-station with a base pressure better than 5$\times$10$^{-11}$mbar. The synchrotron beam-spot size was on the order of 40 $\mu m$.  Measurement temperature was 20K.  Energy resolution was 5 meV for data taken with 20 eV photon energy and 80 meV for data taken with 90 eV photon energy.  The data were collected using a hemispherical Scienta R4000 electron analyser equipped with custom-made deflectors that enable collecting ARPES spectra over a full Brillouin zone without moving the sample. Potassium-dosing experiments were carried out by evaporating potassium in situ from a SAES getter source mounted in the analysis chamber such that dosing is performed  without moving the sample from measurement position. 

\textbf{Crystal growth}: 	High-quality bulk single crystals of 1T’-WTe$_2$ were synthesized by chemical vapor transport (CVT) with bromine as the transport agent. Prior to the CVT growth, a stoichiometric mixture of high purity W (99.9$\%$) and Te (99.9999$+\%$) was heated to 800$^\circ$C for 72 hours in an evacuated quartz tube to form polycrystalline WTe$_2$. For single crystal growth, the pre-compounded polycrystals were ground into a fine powder and transferred into an evacuated 18 cm long, 10 mm inner diameter, 12 mm outer diameter quartz tube together with 3-6 mg/cc of bromine. Using liquid nitrogen, the volatile bromine was condensed with the powder at the bottom end of the ampoule during the quartz sealing process. To minimize oxide based W and Te growth, all sample preparation was done in an argon-filled glovebox to reduce the presence of oxygen and moisture. Additionally, the quartz ampoule was purged and vented with ultrahigh purity argon gas to further reduce the oxygen and moisture content prior to sealing the ampoule. The sealed ampoule was then placed in a four-zone tube furnace and heated up to 900$^\circ$C and 840$^\circ$C at the charges zone and growth zone, respectively, for 6 days. The size of our largest resultant flake was 36 mm$^2$. The x-ray powder diffraction (XRD) pattern of the as grown single crystal is well-matched with the International Centre for Diffraction Powder Diffraction File (ICDD PDF) card 04-007-0799 and confirmed the WTe$_2$ single crystal is in the orthorhombic crystal system, space group Pmn2$_1$. Energy-dispersive x ray (EDX) elemental mapping showed that tungsten and tellurium are distributed evenly, and no other elements were detected except carbon and oxygen, whose presence is strongly believed to be from the vacuum chamber. van der Pauw resistivity measurements suggests that the carrier transport is dominated by electrons with a typical carrier density of 2.77$\times$10$^{20}$ cm$^{-3}$ and specific electrical resistance of about 0.5 m$\Omega$-cm.

\section*{Effect of charge doping on shear displacement in $\gamma$-WTe$_2$ }
\begin{figure}[h]
  \captionsetup{justification=raggedright,width=0.9\columnwidth}
  \includegraphics[width=1\columnwidth]{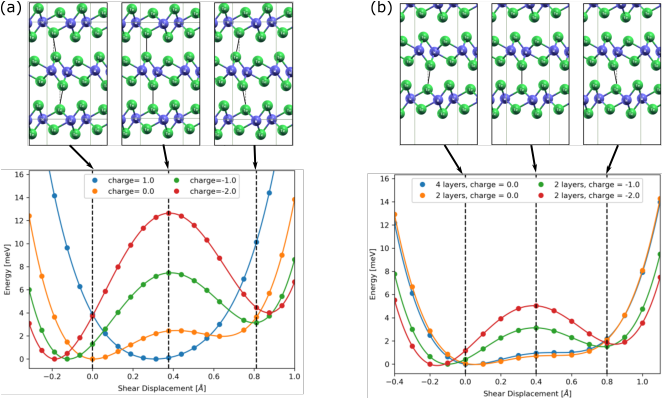}
  \caption{Effect of charge doping on the bulk (a) and slab (b) $\gamma$-WTe$_2$. Charge 1 corresponds to one hole per unit cell, 0 is pristine WTe$_2$, -1 is one electron per unit cell, and -2 is two electrons per unit cell.  Bottom panel shows relative energy as a function of different shear displacements of the middle layer.  Top panels show local crystal structure for select shear displacements of middle plane, with the relative alignment between the nearest Te-Te atoms indicated by black lines between planes. }
  \label{fig:figBulkEnergyProfile}
\end{figure}

Fig. \ref{fig:figBulkEnergyProfile}(a) considers the effect of charge doping on the WTe$_2$ $\gamma$ phase under a bulk geometry, where one of the layers is shifted by various shear displacements. A shear displacement of zero corresponds to the WTe$_2$ $\gamma$ phase. A shear displacement of approximately 0.4\textrm{\AA} corresponds to a geometry where Te atoms on adjacent layers are closest and therefore the effects of antibonding orbitals is locally maximized. Finally a displacement of 0.8\textrm{\AA} causes the Te atoms of adjacent layers to adopt a locally mirrored configuration as compared to the WTe$_2$ $\gamma$ phase. In the absence of charge doping, zero shear displacement (WTe$_2$ $\gamma$) is the most stable state and that there is a small local maximum in configurational energy at 0.4\textrm{\AA} which is the result of the antibonding orbitals between Tellurium atoms of adjacent layers. In the presence of hole doping (charge = 1.0, i.e. one hole per unit cell), the local energy maximum at 0.4\textrm{\AA} vanishes completely and the most stable state has a shear displacement of around 0.3\textrm{\AA}. This is consistent with Ref. \citenum{sie_ultrafast_2019} in which hole-doping accelerates electrons away from antibonding orbitals between adjacent Te atoms, setting adjacent layers in relative oscillatory motion. It should be noted, however, that in the present calculations only one layer is shifted. %It is important to note however that since the effect of the laser is localized, the resulting oscillations of the WTe2 layers is primarily dictated by the non-charged dopped energy profile (charge = 0.0). In Lindenberg's paper, he observes harmonic oscillations for displacements up to around -1.7 to +3.6 pm and begins to see deviations from the harmonic behavior for displacements of around 8 pm which he attributes to a meta-stable state. The meta-stable state that he is referring to is in fact the deviation from a harmonic parabolic potential near 0.4 Angstrom in the non-charged dopped energy profile. Note that Lindenberg's displacements corresponds to shifting both layers in opposite directions by a certain displacement whereas in our calculations only one layer is shifted. To compare to our calculations, Lindenberg's displacements need to be multiplied by a factor of two.

In the presence of electron doping, the local maximum near 0.4\textrm{\AA} is greatly increased. This results from the fact that the additional electronic charge occupies the antibonding orbitals of adjacent Tellurium atoms near the $E_F$. We note that since the repulsion between the Te atoms is increased, the mirror geometry with a shear displacement of 0.8\textrm{\AA} become meta-stable with a local minimum in the energy. However the most stable geometry remains the WTe$_2$ - $\gamma$ state with a displacement of at most -0.2\textrm{\AA}.

%\begin{figure}[h]
 % \captionsetup{justification=raggedright,width=0.9\columnwidth}
 % \includegraphics[width=0.6\columnwidth]{slab_energy_profile}
  %\caption{Effect of charge doping on slab $\gamma$-WTe$_2$. Charge definitions are same as Fig. \ref{fig:figBulkEnergyProfile}.  Bottom panel shows energy relative to minimum as a function of different shear displacements of the bottom layer.  Both 4-layer and 2-layer slabs are shown for charge zero.  Top panels show local crystal structure for select shear displacements of bottom plane, with nearest Te-Te distance indicated by black lines between bottom and middle planes.  }
  %\label{fig:figSlabEnergyProfile}
%\end{figure}

Fig. \ref{fig:figBulkEnergyProfile}(b) shows the effect of charge doping on the WTe$_2$ $\gamma$ phase under a slab geometry with the bottom layer shifted by various shear displacement. The results exhibit a similar qualitative energy profile as compared to the results for the bulk geometry in Fig. \ref{fig:figBulkEnergyProfile}(a). Fig. \ref{fig:figBulkEnergyProfile}(b) shows that in the case of zero charge doping the WTe$_2$ $\gamma$ phase is the global minimum in energy and therefore for energetically most favored state. Compared to the bulk calculations the local maximum at 0.4\textrm{\AA} is approximately half as large. This is due to the fact that under a slab geometry the bottom layer has only half the number of adjacent Tellurium antibonding orbitals due to the presence of the vacuum layer on one side. In the absence of charge doping, we can see that the displacement potential for the bottom layer using a slab geometry with four layers is identical to the case of the slab geometry under a minimum of two layers which indicates that the presence of additional layers to first order does not modify the local potential significantly. In the case of electron doping we only consider the minimum geometry of two layers to ensure that the additional charge is entirely deposited on the two layers. Once again, the presence of electron doping increases the local maximum at 0.4\textrm{\AA}, indicating that the additional charge occupies antibonding orbitals of Tellurium atoms in adjacent layers. The global minimum remains a state very similar to the $\gamma$ phase with a displacement of at most -0.2\textrm{\AA}, and a local minimum in energy is found at the metastable geometry with a shear displacement of 0.8\textrm{\AA}.

\begin{figure}[h]
  \captionsetup{justification=raggedright,width=.9\columnwidth}
  \includegraphics[width=1\columnwidth]{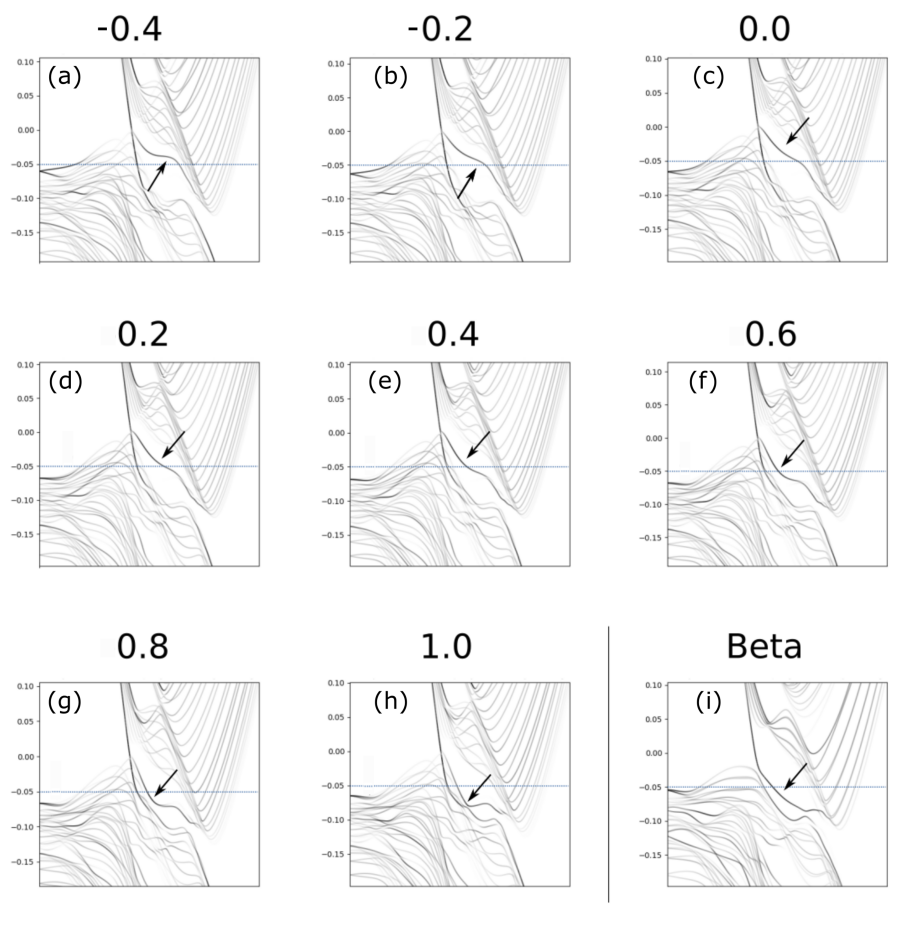}
  \caption{Evolution of low-energy electronic structure as a function of shear displacement of the top layer with 'face A' termination. Labels above (a)-(h) indicate amount of shear displacement of top layer in \textrm{\AA}, and (i) shows $\beta$ phase for comparison. Arrows point to surface states of interest.}
  \label{fig:figFaceBShifts}
\end{figure}

\begin{figure}[h]
  \captionsetup{justification=raggedright,width=.9\columnwidth}
  \includegraphics[width=1\columnwidth]{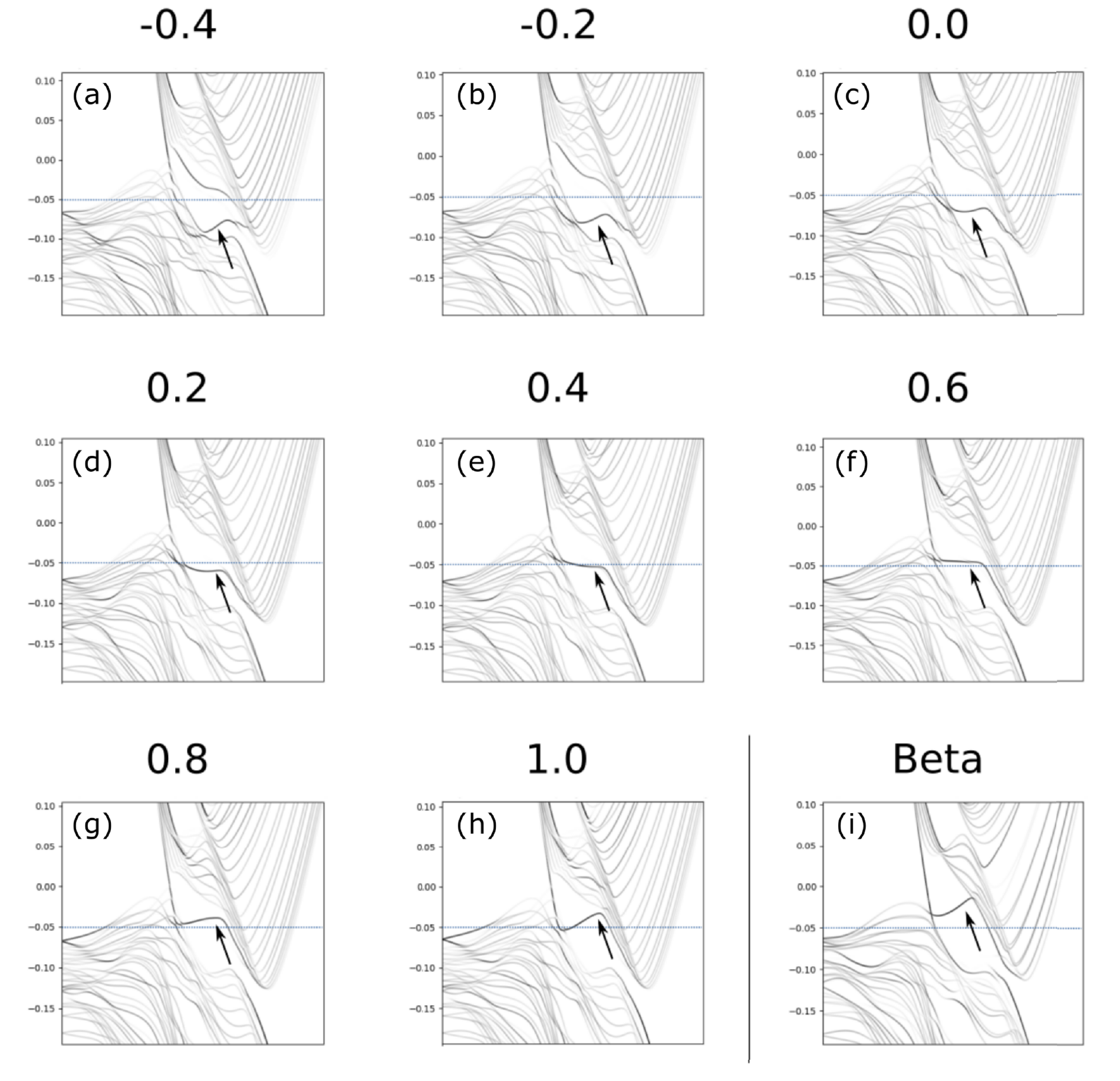}
  \caption{Evolution of low-energy electronic structure as a function of shear displacement of the top layer with 'face B' termination. Labels above (a)-(h) indicate amount of shear displacement of top layer in \textrm{\AA}, and (i) shows $\beta$ phase for comparison. Arrows point to surface states of interest.}
  \label{fig:figFaceAShifts}
\end{figure}

In both of the above calculations the $\gamma$ phase remains the most stable state with a shear displacement of at most -0.2 \textrm{\AA}. However there are several limitations in the above calculations that prevent determination of energy changes down to several meV, such that the possibility of the meta-stable state (shear displacement of 0.8\textrm{\AA}) being a global minimum in energy cannot be excluded. First the van der Waals interaction was modeled empirically using DFT+D3 which does not capture effects due to electron density and hence is fundamentally incapable of modeling the effect of a change in charge density. The second limitation in the case of the slab calculations is that only a minimal of two layers were considered to ensure that the added charge is fully localized on only two layers in a consistent fashion. In light of these limitations, the possibility that the meta-stable state in fact become the stable state cannot be excluded. However, the above calculations illustrate that the addition or removal of charge modifies the strength of antibonding orbitals between Tellurium atoms in adjacent layers suggesting a mechanism for generating shear displacements using electron doping.

Fig. \ref{fig:figFaceBShifts} systematically shows the evolution of the 'face A' surface state as a function of various displacements of the surface layer from -0.4 to 1.0 \textrm{\AA}. Displacements outside of this range are unlikely, due to the high energy barriers at -0.4 and 1.0 \textrm{\AA} as illustrated Fig. \ref{fig:figBulkEnergyProfile}(b).  It is seen that the surface state between the electron and hole pockets, marked by an arrow, evolves from being below $E_F$ to being above $E_F$.  Notably the meta-stable configuration reproduces the key experimental observable that this band is longer visible in ARPES experiments that only measure occupied states.  Also note the qualitative similarity between the metastable configuration (0.8 \textrm{\AA} shear displacement) and the $\beta$ phase.

Fig. \ref{fig:figFaceAShifts} systematically shows the evolution of the 'face B' surface state as a function of various displacements of the surface layer from -0.4 to 1.0 \textrm{\AA}.  Again, the meta-stable configuration reproduces the key experimental observable that a surface band between the electron and hole pocket appears in ARPES spectrum, and also has qualitative similarity to the $\beta$ phase.

\begin{figure}[h]
  \captionsetup{justification=raggedright,width=0.9\columnwidth}
  \includegraphics[width=1\columnwidth]{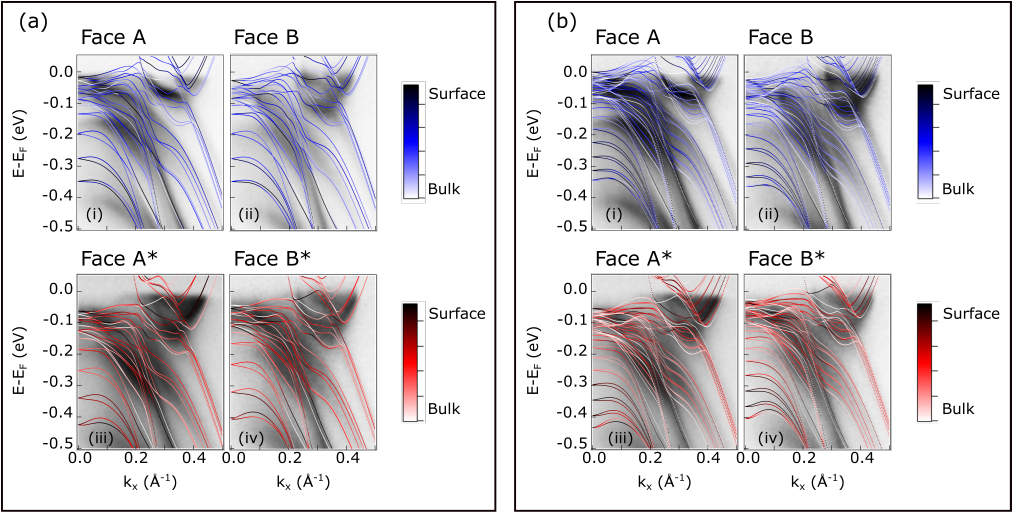}
  \caption{Superposition of DFT calculated bands (colored, 3 unit cells (a) and 6 unit cells (b)) with measured band structure along $\Gamma - Y$ direction. (i-ii) Band structure from Face A and B, respectively. (iii-iv) Band structure after doping induced transition from Face A* and B*, respectively}
  \label{fig:figS2}
\end{figure}

%\begin{figure}[h]
 % \captionsetup{justification=raggedright,width=0.9\columnwidth}
 % \includegraphics[width=0.65\columnwidth]{Figsixslab.png}
 % \caption{Superposition of DFT calculated bands (colored, 6 unit cells) with measured band structure along $\Gamma - Y$ direction. (a-b) Band structure from Face A and B, respectively. (c-d) Band structure after doping induced transition from Face A* and B*, respectively}
 % \label{fig:figS2b}
%\end{figure}

Fig. \ref{fig:figS2} shows the same data and calculations as in Fig. 1 of the main text, except with DFT Slab calculation of the $\beta$ phase (3 unit cells ,6 layers (a)) overlaid on the ARPES data to qualify agreement, particularly with regards to evolving surface states. A calculation taking into account a thicker slab (6 unit cells) is shown in Fig. \ref{fig:figS2}(b). Both agree with the general trend reported in the manuscript, justifying the use of the 3 unit cell system.

%\begin{figure}[h]
 % \captionsetup{justification=raggedright,width=0.9\columnwidth}
 % \includegraphics[width=0.65\columnwidth]{Difference_derivative.png}
 % \caption{Difference of second derivative spectra before and after dosing for face A and face B.  Second derivative spectra after dosing were shifted by 30meV to reflect shifting of chemical potential.}
  %\label{fig:figS_2ndDeriv}
%\end{figure}

We now turn briefly to the K-coverage for the first phase transition.  As we will discuss in the next section, the experimentally-controlled parameter during alkali metal dosing is the dosing time, typically broken down into equal cycles.  However, mapping the dosing time or cycle number onto an electron count is problematic because the probability that deposited K will ionize and donate electrons to the host is itself a function of time or cycle number.  Instead, the more rigorous experimental measure is the shift in chemical potential, which can be mapped onto the number of donated electrons with knowledge of the band structure and assumptions about how much the doped electron delocalizes into the bulk.  For a chemical potential shift of 0.03 eV, applicable to the first phase transition, we get the following values for the number of doped electrons:
\begin{itemize}
  \item Doped electrons delocalized over entire bulk: 0.03 electrons per unit cell
  \item Three unit cell slab with charge projected over: \begin{itemize}
      \item top unit cell: 0.03 electrons per unit cell
      \item central unit cell: 0.039 electrons per unit cell
      \item bottom unit cell: 0.031 electrons per unit cell
  \end{itemize}
\end{itemize}
Thus, the first phase transition corresponds to a doping value of 3-4$\%$.

\section*{Second phase transition: additional data}
The second phase transition corresponds to a chemical potential shift of $\approx 100-130 meV$, which is equivalent to 11-16$\%$ doping delocalized over the bulk.  

\begin{figure}[h]
  \captionsetup{justification=raggedright,width=0.7\columnwidth}
  \includegraphics[width=0.6\columnwidth]{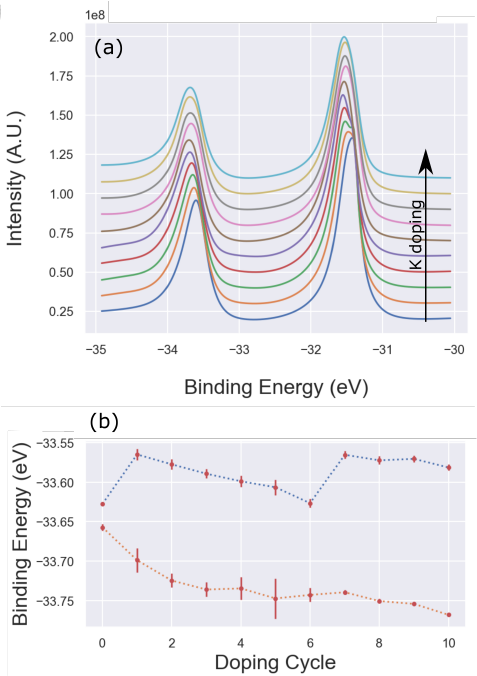}
  \caption{(a) W $4\text{f}\frac{5}{2}$ and $4\text{f}\frac{7}{2}$ core level at different doping steps. (b) W core level binding energies, fit from panel (a).}
  \label{fig:figS4}
\end{figure}

The second phase transition is apparent in both W and K core levels, measured by x-ray photoemission spectroscopy (XPS).  The XPS data in Fig.  \ref{fig:figS4}(a) show W core levels.  These data were fit with Voigt functions to track the position of the peaks, and the evolution as a function of doping cycle is shown in Fig. \ref{fig:figS4}(b).  In all core levels, an initial jump in peak position is observed after the first dosing cycle, due to the sudden change of overall chemical environment. With subsequent cycles, the binding energy downshifts due to the overall doping mechanism that involves the creation of a dipole field at the interface. However, when the second transition occurs (cycle 7), the W core level corresponding to the top W layer in the unit cell stops evolving monotonically and jumps to higher binding energy. Over subsequent cycles, the peak position again evolves monotonically.

%\begin{figure}[h]
 % \captionsetup{justification=raggedright,width=0.9\columnwidth}
  %\includegraphics[width=0.6\columnwidth]{FigS11_v2.png}
  %\caption{W core level binding energies, fit from Fig. %\ref{fig:figS4} with guides-to-the-eye.}
  %\label{fig:figS3}
%\end{figure}

\begin{figure}[h]
  \captionsetup{justification=raggedright,width=0.75\columnwidth}
  \includegraphics[width=0.8\columnwidth]{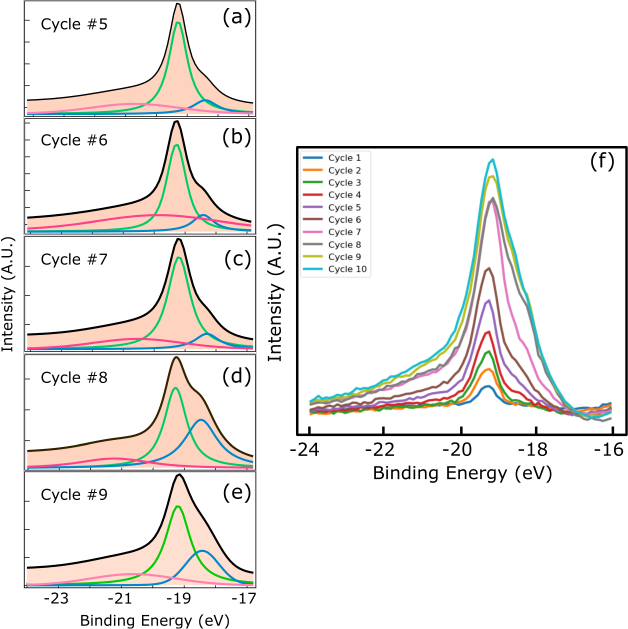}
  \caption{a-e) K $3p$ core level at different doping cycles. The peak is fit with two Voigt functions (green, blue) and a a broad Gaussian. A Shirley function for the background is also used (not displayed). (f) K $M_{3}-3p_{3/2}$ core level at different doping cycles. }.
  \label{fig:figS6}
\end{figure}

The K core levels (Fig. \ref{fig:figS6}) also show a change at the same doping level.  Both before and after the transition, two peaks are observed.  The one at lower binding energy ($\approx -18.4 eV$ in the present study) is consistent with the 3p core level in potassium metal, as measured and tabulated by earlier studies\cite{Petersson:Kmetal_KPS,Cardona:CoreLevels}.  The one at slightly higher binding energy ($\approx -19.3 eV$ in the present study) is interpreted as ionized potassium which has donated its electron to the substrate.  Previous studies of potassium deposited onto metal surfaces have demonstrated a shift of the K 3p core level from slightly higher binding energy to the bulk value as monolayer coverage is approached\cite{Broden:Kdosing_Fe,Ozawa:K_adsorb}.  The adsorption of alkali metals on a substrate is interpreted as initially being ionic (donating electrons to the substrate) and becoming more neutral (bonding among adatoms) with further dosing\cite{Trioni:AlkaliAdsorption_review}.  The observed evolution of K core levels in Fig. \ref{fig:figS6}(a-e) is consistent with this picture.  The rise of the peak corresponding to neutral K after the second transition is consistent with our interpretation of bonding within the K layer being one ingredient of that transition.  K core levels for all dosing cycles are plotted in Fig. \ref{fig:figS6}(f).  A reduction in peak intensity is \textit{not} observed after the second phase transition, indicating that intercalation is not likely.  %The one at higher binding energy corresponds to isolated K atoms while the one at lower binding energy corresponds to clusters of K atoms.  At the transition, the second peak suddenly becomes more intense relative to the first one, indicating a predominance of clustered or mutually interacting K atoms.  This together with first principles calculations provides a rationale for interpreting the second phase transition in terms of a 2D metal forming and starting to interact with WTe$_2$.

%\begin{figure}[h]
 % \captionsetup{justification=raggedright,width=0.8\columnwidth}
 % \includegraphics[width=0.4\columnwidth]{FigS11.png}
 % \caption{(a) K $M_{3}-3p_{3/2}$ core level at different doping cycles.}
 % \label{fig:figS11}
%\end{figure}

\begin{figure}[h]
  \captionsetup{justification=raggedright,width=1\columnwidth}
  \includegraphics[width=0.7\columnwidth]{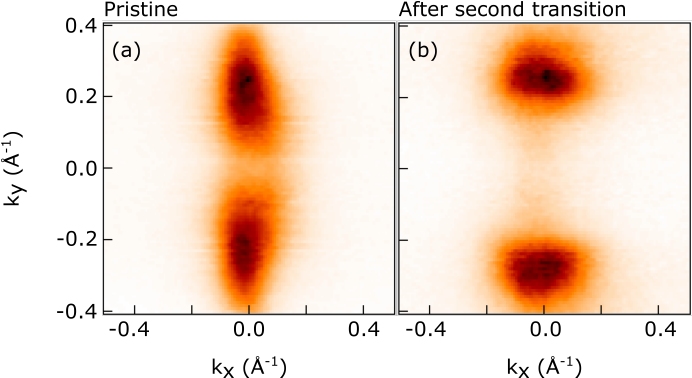}
  \caption{Constant energy contours collected at 91 eV photon energy, for pristine crystal (left, $E=E_F$) and after the second transition (right, $E=-130 meV$).  }
  \label{fig:figS7}
\end{figure}

The second phase transition is also manifested in the band structure beyond what is discussed in the main text.  One example is constant energy contours which change orientation, as shown in Fig. \ref{fig:figS7}  These contours are taken at \textit{equivalent energies in the band structure} ($E_F$ and $-130 meV$) before dosing and after the second phase transition.  A clear change in the shape of the contours is observed.  Initially Fermi surfaces are elongated along the $k_y$ direction, reflecting the quasi-1D tendency of the crystal structure induced by the W zigzag chains.  Afterward, the contours are elongated along $k_x$.  This change cannot simply be attributed to a Lifshitz transition due to a changing chemical potential, because the constant energy contour after the second phase transition are not taken at the chemical potential.  Instead, it reflects a large rearrangement of the band structure.

\begin{figure}[h]
  \captionsetup{justification=raggedright,width=1.0\columnwidth}
  \includegraphics[width=1\columnwidth]{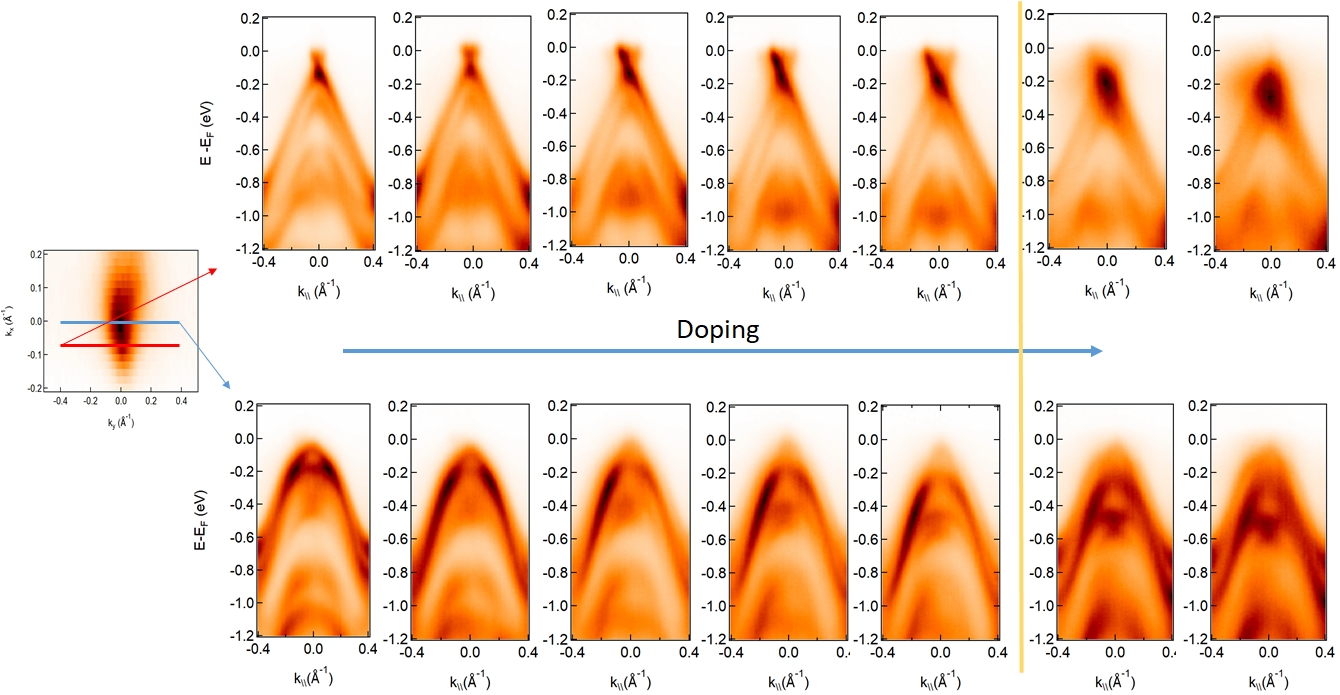}
  \caption{Evolution of two off-high-symmetry cuts as a function of dosing across second phase transition.  Cut positions are indicated in constant energy contour on left. Top row: cut through center of electron pocket, showing additional panels from what is shown in main text. Bottom: cut through hole pocket taken at same dosing amounts. Vertical yellow line marks  second phase transition.}
  \label{fig:figS8}
\end{figure}

Fig. \ref{fig:figS8} shows how two off-high-symmetry cuts evolve with K dosing across the second phase transition.  The first cut (top row) is through the center of the electron pocket. Previous ARPES studies \cite{Thirupathaiah:TempIndeptBandStructWTe2} have identified the observed Dirac-like dispersion as a surface state. Initial dosing trivially shifts the chemical potential and slightly broadens the band, but keeps the dispersion intact.  After the second threshold is reached, the upper branch of the dispersion is no longer observed, and the distinction between this effect and the trivial broadening observed in previous dosing cycles is clear.  The second off-high-symmetry cut (bottom row) goes through the hole pocket, and is initially characterized by two hole-like dispersions near $E_F$ which can clearly be distinguished from one another.  Across the second phase transition, the one at lower binding energy disappears and an additional feature appears at $E\approx-0.5 eV$.

\begin{figure}[h]
  \captionsetup{justification=raggedright,width=1\columnwidth}
  \includegraphics[width=1\columnwidth]{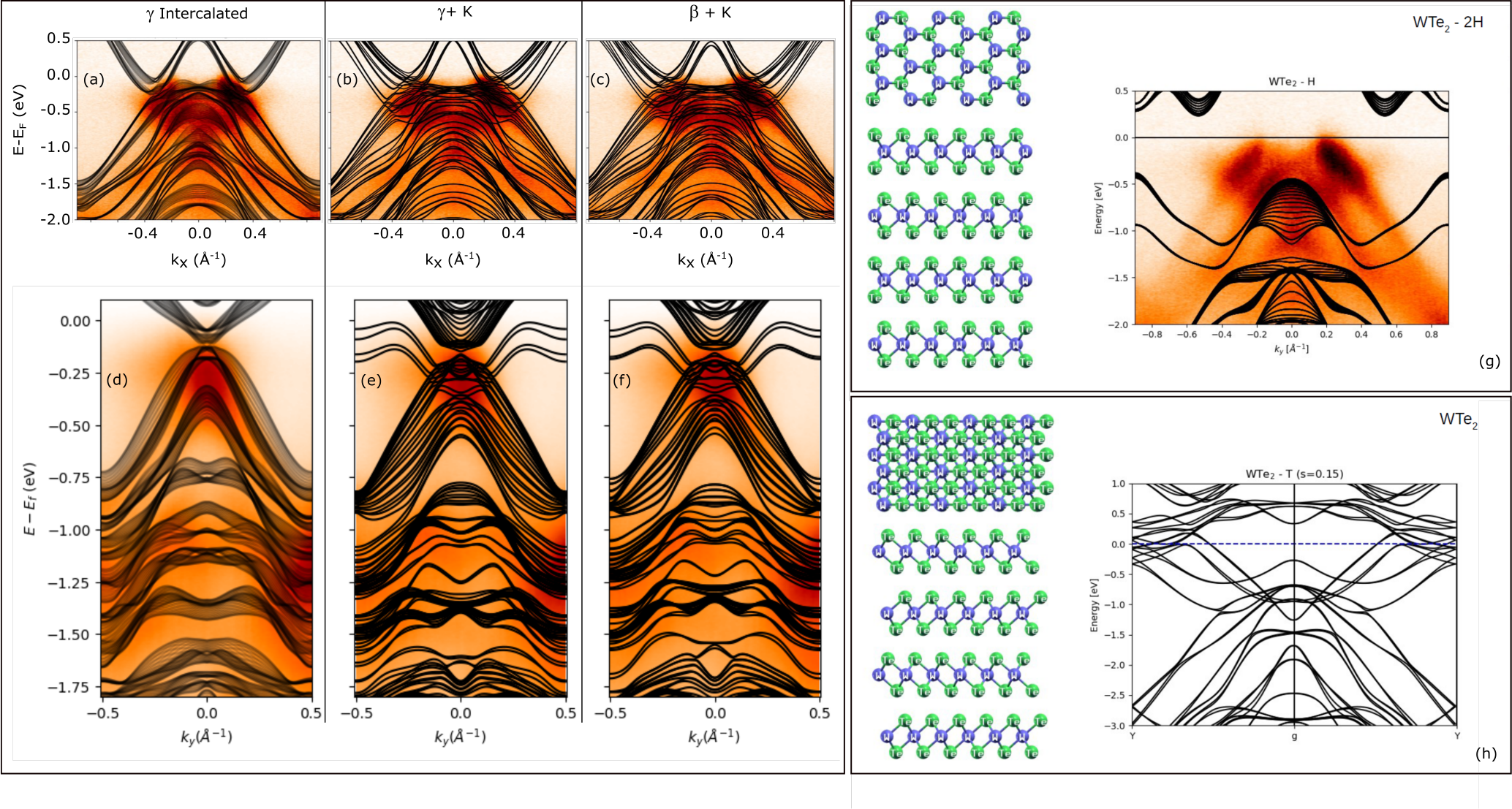}
  \caption{DFT calculation ovelayed on ARPES data considering intercalation (a,d), K-adlayer hybridizing with $\gamma$ (b, e), and K-adlayer hybridizing with $\beta$ (c,f).  (a)-(c) are high-symmetry $\Gamma-Y$ cut and (d)-(f) are off-high-symmetry cut used in top row of Fig. \ref{fig:figS8}. Other bulk band structure calculations.  (g): Crystal structure and calculation for semiconducting hexagonal 2H phase.  (h): Crystal structure and calculation for undistorted T-phase. }
  \label{fig:band_overlay}
\end{figure}

%\begin{figure}[h]
%  \captionsetup{justification=raggedright,width=1\columnwidth}
%  \includegraphics[width=1\columnwidth]{FigS12.png}
%  \caption{Other bulk band structure calculations.  Left: Crystal structure and calculation for semiconducting hexagonal 2H phase.  Right: Crystal structure and calculation for undistorted T-phase.} 
%  \label{fig:figS12}
%\end{figure}

Fig. \ref{fig:band_overlay}(a-f) shows a comparison of WTe$_2$ intercalated with K, K/$\gamma$-WTe$_2$ surface hybridization, and K/$\gamma$-WTe$_2$ surface hybridization. In all cases, there is one K-atom per unit cell contributing 0.5 electrons. The hybridized band structure is very similar considering the $\gamma$ or $\beta$ phase of WTe$_2$, notably reproducing the extra spectral weight away from the main bands in high-symmetry and off-high-symmetry cuts.  

Other scenarios that were considered are shown in Fig. \ref{fig:band_overlay}(g,h).  The 2H phase is metastable and  reached in some parameter regimes.  For example, electrostatic doping in MoTe$_2$ induces a structural phase transition at room temperature which was identified as a transition from the 1T' phase to the 2H phase\cite{Wang:MonolayerMoTe2PhaseChange}.  However, this band structure does not have good agreement with what is observed after the second phase transition. The T phase (undistorted octahedral) is not stable and also has poor agreement with the data.

\begin{figure}[h]
  \captionsetup{justification=raggedright,width=1\columnwidth}
  \includegraphics[width=0.5\columnwidth]{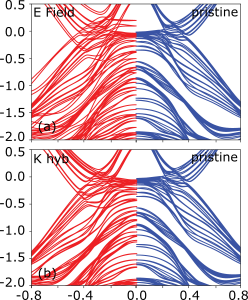}
  \caption{Comparison along high symmetry ($\Gamma-Y$) cut for slab calculation considering electric field of 10 GV/m (top) and hybridization with K-overlayer  with 1 potassium per unit cell, donating 0.5 electrons each (bottom).  Blue bands on left are pristine $\gamma$ phase and right bands (left) are for aforementioned perturbations.  }
  \label{fig:figS_EH}
\end{figure}

In the main text, we show the effects of both hybridization (one K atom per unit cell, 0.5 electrons per K) and electric field (10 GV/m) at the $\Gamma$ point, and Figure \ref{fig:figS_EH} shows the the full $\Gamma-Y$ cut for a three-unit-cell slab.  The values of doping and electric field are chosen to reflect an upper bound to emphasize the effect of the perturbation.  For example, the doping corresponding to our chemical potential shift is $\approx 0.11-0.16$ and an electric dipole with charges separated by the height of a (half) unit cell corresponds to (14.5) 3.7 GV/m.

\begin{figure}[h]
  \captionsetup{justification=raggedright,width=1\columnwidth}
  \includegraphics[width=0.8\columnwidth]{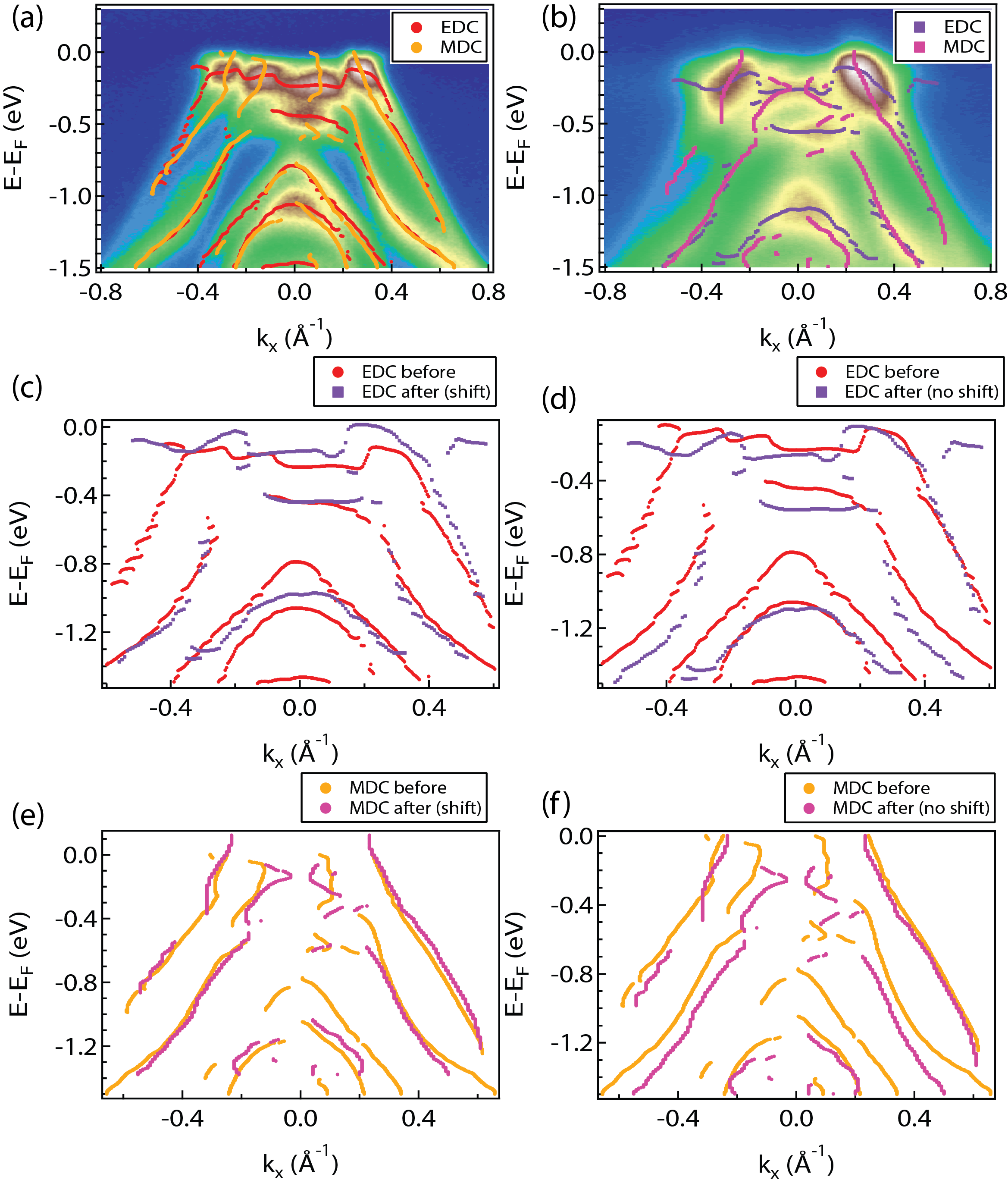}
  \caption{Comparison along high symmetry ($\Gamma-Y$) before and after second phase transition. (a) Before second phase transition, with EDC and MDC peak position overlays. (b) After second phase transition, with EDC and MDC peak position overlays. (c) EDC peak positions before and after second phase transition, with 'after' data shifted by 0.12 eV. (d) EDC peak position before and after phase transition without energy shifting. (e) same as (c), but for MDC peaks. (f) same as (d), but for MDC peaks.}
  \label{fig:figS_EDC_MDC}
\end{figure}

Figure \ref{fig:figS_EDC_MDC} compares EDC and MDC peak positions before and after the second phase transition to describe how the chemical potential shift is attained.  Peak positions were obtained after smoothing followed by a first derivative, and as such, shoulder features are not captured.  The bands at the $\Gamma$ point do not shift monotonically across the second phase transition, so there is no choice in chemical potential shift that causes them to coincide.  However, the bands further from $\Gamma$ appear to be less strongly affected and can be aligned with a chemical potential shift 0.1-0.13 eV.

\section*{First-Principles calculations: lattice parameters and type-II Weyl semimetal}
\begin{figure}[h]
  \captionsetup{justification=raggedright,width=1.0\columnwidth}
  \includegraphics[width=1\columnwidth]{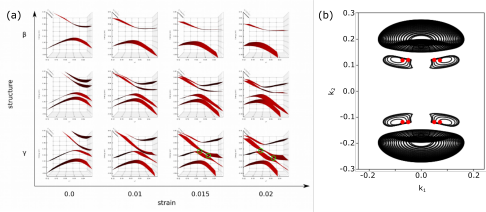}
  \caption{(a) Effect of c-axis compression on Weyl points and topological phase transition.  Each column shows the evolution from the $\gamma$ to the $\beta$ phase (bottom to top), similar to Fig. 3 in main text.  Each column differs by amount of c-axis compression.  Sufficient compression supports Weyl points in the $\gamma$ phase. (b) Calculated positions of Weyl points (red) in the Brillouin zone.  Axes $k_1$ and $k_2$ are scaled such that $\pm1$ is the Brillouin zone boundary.}
  \label{fig:figS9}
\end{figure}

Non-relaxed experimental crystal parameters were used to determine all layered
WTe$_2$ structures as it was found that geometric relaxation worsened the
agreement with ARPES measurements. As noted in previous
studies~\cite{OriginsTransitions}, empirical van der Waals correction schemes
such as DFT-D2 and DFT-D3 are not able to accurately model the interlayer
interaction and therefore lead to inaccurate layer separations.
Several experimental parameters have been reported for the WTe$_2$-$\gamma$
phase of which two sets of parameters~\cite{WTe2Brown,WTe2_gamma2} are widely
referenced through the literature and were taken at different temperatures.
Consistent with previous studies~\cite{TunableWeylMoWTe2}, we find that the
room temperature crystal structure for WTe$_2$-$\gamma$ reported by
Brown~\cite{WTe2Brown} does not exhibit Weyl points although the system is
remarkably close to realizing them.
Although not the focus of the manuscript, the type-II Weyl semimetal phase, which is only supported by the $\gamma$ crystal structure was also studied.  It was found that approximately a 1.5\% compression along the c-axis of the
unit cell is required for the Weyl points to emerge (see Fig. \ref{fig:figS9}). The effect of uniaxial
compression is consistent with several previous works which have reported that
compression along the c-axis increases the Weyl point separation, thereby
stabilizing the Weyl-semimetal state~\cite{WTe2Bernevig,OriginsTransitions}.  This sensitivity to c-axis parameters makes the bulk Weyl points of WTe$_2$ even more elusive for ARPES experiments, as they may already be annihilated at temperatures high enough to thermally populate their location above $E_F$.

%\begin{figure}[h]
%  \captionsetup{justification=raggedright,width=0.75\columnwidth}
%  \includegraphics[width=0.25\columnwidth]{FigS1.png}
%  \caption{Calculated positions of Weyl points (red) in the Brillouin %zone.  Axes $k_1$ and $k_2$ are scaled such that $\pm1$ is the %Brillouin zone boundary.}
%  \label{fig:figS1}
%\end{figure}
In contrast with results reported in \cite{WTe2Bernevig}, we find that the
experimental parameters measured at a lower temperature\cite{WTe2_gamma2},
although slightly closer to realizing Weyl point crossings, exhibits only
small changes in the band structure. This result was further confirmed using
the all-electron full-potential linear augmented plane wave method (FP-LAPW)
as implemented in ELK\cite{ELK}. It should be noted that\cite{WTe2Bernevig}
reported exceptionally small Weyl point separations, far beyond spectroscopic
experimental resolution.
Consequently, taking into consideration the similarity between the band
structures of the two crystal parameters, in this study we have use the
crystal structure for WTe$_2$-$\gamma$ reported by Brown \cite{WTe2Brown} in
all calculations as it is more commonly used though out the literature and
enables for a more direct comparison to the structures of MoTe$_2$-$\gamma$
and MoTe$_2$-$\beta$ reported in the same article. An approximate crystal
structure for the WTe$_2$-$\beta$ phase was obtained by replacing Mo with W in
the experimental structure for MoTe$_2$-$\beta$~\cite{WTe2Brown} and rescaling the lattice
constants to be consistent with those of WTe$_2$-$\gamma$.

The positions of the Weyl points in the Brillouin zone for 1.5\% compression along the c-axis is shown in Fig. \ref{fig:figS9}(b).  As acknowledged in previous literature, Weyl points of opposite chirality are very close to each other, making it very difficult to observe Fermi arcs connecting them.  Moreover, the bulk bands for different c-axis lattice constants (and also for the $\gamma$ vs $\beta$ phase) are almost indistinguishable, making the type-II Weyl semimetal phase in WTe$_2$ very difficult to confirm with ARPES.

%The $\gamma$ to $\beta$ transition can be visualized as a shear displacement of a unit cell relative to the one below, and Fig. \ref{fig:figS10} demonstrates how, upon electron doping, the local minimum energy is realized for a small shear displacement of the top unit cell, not for the undistorted structure.

%\begin{figure}[H]
  %\captionsetup{justification=raggedright,width=0.7\columnwidth}
  %\includegraphics[width=0.7\columnwidth]{FigS10.png}
  %\caption{Surface shear displacement. Left: crystal structure of $\gamma$ phase. Right: energy associated with shear displacement of top layer, for different amounts of electron doping. Vertical dashed lines mark local minima.} 
  %\label{fig:figS10}
%\end{figure}

\bibliography{Final_WTe2_SI_PhysRev_v6.bib}